\documentclass[journal]{IEEEtran}
\IEEEoverridecommandlockouts
\usepackage{cite}
\usepackage{amsthm}
\usepackage{amsmath}
\usepackage{comment}
\usepackage{booktabs}
\usepackage{stfloats}
\usepackage{bm}
\usepackage{setspace}
\usepackage{graphicx}
\usepackage{subfigure}
\usepackage[usenames,dvipsnames]{color}
\usepackage{amsfonts,amssymb,amsmath}
\usepackage{subfigure}
\usepackage{enumerate}
\usepackage{cases}
\usepackage{array}
\usepackage{colortbl}
\usepackage{multirow}
\usepackage{arydshln}
\usepackage{epstopdf}
\usepackage{threeparttable}
\usepackage{caption}
\usepackage{diagbox}
\usepackage{xcolor}
\usepackage{xpatch}

\makeatletter

\newcommand{\Rmnum}[1]{\expandafter\@slowromancap\romannumeral #1@}

\makeatother

\definecolor{Gray}{gray}{0.85}
\definecolor{mycyan}{cmyk}{.3,0,0,0}

\newcolumntype{a}{>{\columncolor{Gray}}c}
\newcolumntype{b}{>{\columncolor{white}}c}
\newcolumntype{d}{>{\columncolor{mycyan}}c}

\allowdisplaybreaks
\def\BibTeX{{\rm B\kern-.05em{\sc i\kern-.025em b}\kern-.08em
    T\kern-.1667em\lower.7ex\hbox{E}\kern-.125emX}}

\begin{document}
\title{Timeliness of Status Update System: The Effect of Parallel Transmission Using Heterogeneous Updating Devices
}

\author{\IEEEauthorblockN{Zhengchuan~Chen,~\IEEEmembership{Senior Member,~IEEE}, Kang~Lang, \\
Nikolaos Pappas,~\IEEEmembership{Senior Member,~IEEE}, Howard H. Yang,~\IEEEmembership{Member,~IEEE}, \\
Min Wang,~\IEEEmembership{Member,~IEEE}, Zhong~Tian,~\IEEEmembership{Member,~IEEE}, and Tony Q. S. Quek,~\IEEEmembership{Fellow,~IEEE}}

\thanks{This paper was presented in part at the IEEE ICC 2023 \cite{early_work-conf}.

Z. Chen, K. Lang, and Z. Tian are with the School of  Microelectronics and Communication Engineering, Chongqing University, Chongqing, China. (E-mails:~\{czc,~kangl,~ztian\}@cqu.edu.cn).

N. Pappas is with the Department of Computer and Information Science, Link\"{o}ping University, 581 83 Link\"{o}ping, Sweden (E-mail: nikolaos.pappas@liu.se).

H. H. Yang is with the Zhejiang University/University of Illinois at Urbana-Champaign Institute, Zhejiang University, Haining 314400, China. (E-mail: haoyang@intl.zju.edu.cn).

M. Wang is with the School of Optoelectronics Engineering, Chongqing University of Posts and Telecommunications, Chongqing, China, and is also with Postdoctoral Research Center of Chongqing Key Laboratory of Optoelectronic Information Sensing and Transmission Technology, Chongqing University of Posts and Telecommunications, Chongqing 400065, China. (E-mail:~wangm@cqupt.edu.cn).

T. Q. S. Quek is with the Pillar of Information Systems Technology and Design, Singapore University of Technology and Design, Singapore. (E-mail: tonyquek@sutd.edu.sg).
}
}

\maketitle

\begin{abstract}
  Timely status updating is the premise of emerging interaction-based applications in the Internet of Things (IoT). Using redundant devices to update the status of interest is a promising method to improve the timeliness of information. However, parallel status updating leads to out-of-order arrivals at the monitor, significantly challenging timeliness analysis. This work studies the Age of Information (AoI) of a multi-queue status update system where multiple devices monitor the same physical process. Specifically, two systems are considered: the \emph{Basic System}, which only has type-1 devices that are ad hoc devices located close to the source, and the \emph{Hybrid System}, which contains additional type-2 devices that are infrastructure-based devices located in fixed points compared to the \emph{Basic System}. Using the Stochastic Hybrid Systems (SHS) framework, a mathematical model that combines discrete and continuous dynamics, we derive the expressions of the average AoI of the considered two systems in closed form. Numerical results verify the accuracy of the analysis. It is shown that when the number and parameters of the type-1 devices/type-2 devices are fixed, the logarithm of average AoI will linearly decrease with the logarithm of the total arrival rate of type-2 devices or that of the number of type-1 devices under specific condition.
  It has also been demonstrated that the proposed systems can significantly outperform the FCFS M/M/$N$ status update system.
\end{abstract}


\section{Introduction}\label{sec_1}
\begin{figure*}[t]
  \centering
  \includegraphics[scale=0.5]{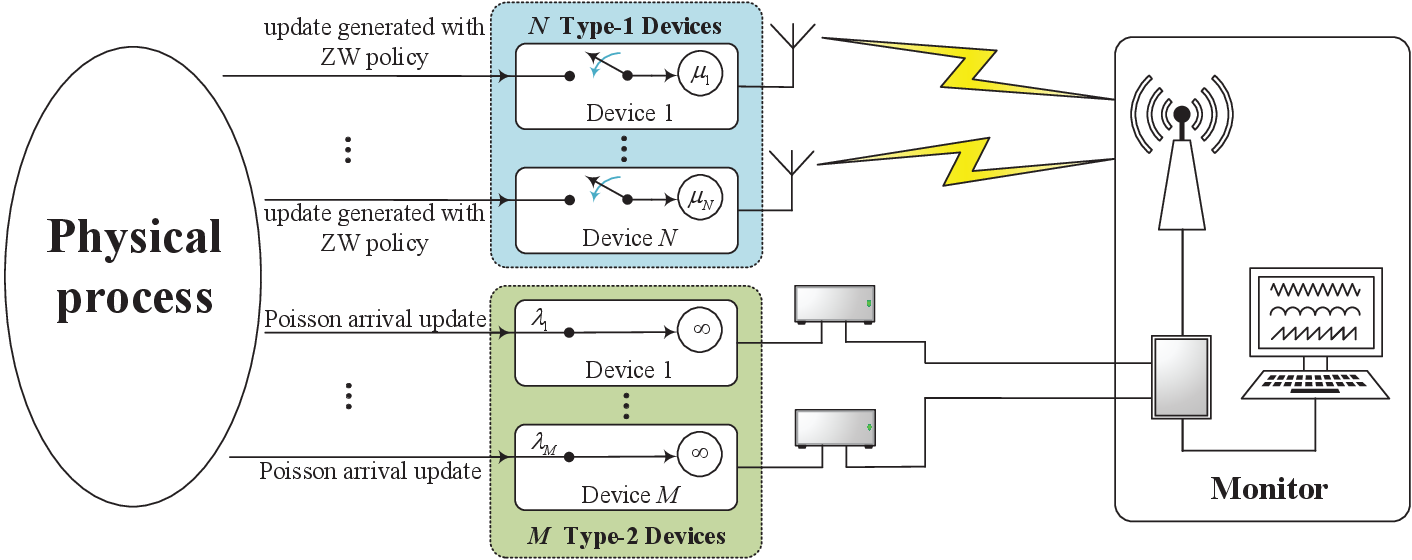}
  \caption{A schematic view of multi-queue status update system, where type-1 device has a sampler and type-2 device does not.}\label{SystemModel}
  \vspace{-1em}
\end{figure*}

A broad range of Internet of Things (IoT) applications have stringent timeliness requirements. For instance, instant information about temperature and humidity from forests can contribute to forest fire prediction and prevention\cite{queue_initial_work}. In vehicular networks, timely status updates are critical to ensure the real-time and accurate control of the vehicle\cite{motivation_vehicle3, motivation_vehicle2}. In remote surgery, doctors' operations (e.g., the changes in the position of surgical tools) need to be executed in time to ensure the safety of surgery\cite{motivation_remote-surgery}. In these applications, the freshness of information directly affects the value of applications and even the safety of people.

The Age of Information (AoI) has been proposed as a pertinent metric for measuring the timeliness of information. AoI is the time elapsed since the latest update received at the destination was generated\cite{queue_initial_work,AoI_book_NP1,AoI_book_NP2,AoI_survey}.
In addition, AoI is an instantaneous concept, and the most commonly used metric of AoI is the time average of AoI, i.e., average AoI\cite{AoI_book_NP1}. Another widely used metric of AoI is the Peak AoI (PAoI), which describes the maximum instantaneous AoI before the next update is accepted\cite{queue_schedule_packet-management_journal}.

The classical works on AoI mainly focused on analyzing the AoI performance under elementary queueing disciplines (single-source single-server queues) and scheduling strategies that can improve the AoI performance (e.g. service disciplines and packet management strategies) \cite{queue_schedule_packet-management_journal, queue_schedule_LCFS, queue_schedule_deadline, np_schedulingTCOM23, np_packet-management, np_optimal-policies, np_distribution}. In \cite{queue_initial_work}, the authors derived the average AoI of the M/M/1, M/D/1, and D/M/1 queueing systems under the first-come, first-served (FCFS) discipline. The average AoI of the M/M/1 queueing system with and without preemption under last-come-first-served (LCFS) discipline was derived in \cite{queue_schedule_LCFS}. The results showed that the LCFS discipline outperforms the FCFS discipline since the new update would be served before the old one. In \cite{queue_schedule_packet-management_journal}, the author considered several lossy queues using package management strategies and found that dropping redundant packets when the arrival rate of data is large and packet replacement in the buffer leads to a small average AoI. Besides, the deadline strategy was shown to be able to reduce AoI\cite{queue_schedule_deadline}. Considering the general distribution of queue arrival and service processes, the stationary distribution of the AoI of the GI/GI/1 queues under four service disciplines was also presented\cite{queue_general-distribution_distribution}. In \cite{queue_general-distribution_AoI}, the authors investigated the average AoI of the G/G/1/1 systems under two service disciplines.

Another line of research on elementary queues is the multi-source setup.
The average AoI of a multi-source M/M/1 queue was first studied under the FCFS discipline \cite{multi-source_init} and then revisited to correct an error \cite{multi-source_revisited}. For the multi-source M/M/1 queue, it was shown that the LCFS discipline also performs better than the FCFS discipline in terms of the average AoI \cite{multi-source_SHS-propose}. The study of the average AoI of the multi-source system has been extended to the multi-source M/G/1 queue, where three approximate results of the per-source average AoI were established \cite{multi-source_MG1-approximate}. For the multi-source M/G/1/1 non-preemptive system, its average AoI and the optimization of the average AoI were investigated in \cite{multi-source_schedule_DDP}.
Note that to deal with the increased system complexity caused by the coupling of multiple sources, an effective approach based on Stochastic Hybrid Systems (SHS) was proposed in \cite{multi-source_SHS-propose} and has been widely used\cite{multi-source_revisited,multi-source_SHS-propose,multi-server_MMc}.
Based on SHS, finding the average AoI is equivalent to seeking the stationary distribution of a finite-state Markov chain.

The aforementioned works are mainly based on single-server queues.
However, the system with only one server cannot grasp the feature of the disorderly arrival of packets in the network since the packets can be delivered through multiple paths\cite{multi-server_MM2_journal}. Thus, the scenario that multiple servers work for a single queue was considered. Two special cases, the M/M/2 and M/M/$\infty$ queues, were considered in \cite{multi-server_MM2_journal}. The results showed that the M/M/2 queue can reduce the AoI by about half compared with the M/M/1 queue. For infinite-server queues, the author in \cite{multi-server_infinite} considered general arrival interval distribution/service time distribution and derived the probability distribution of AoI in a GI/GI/$\infty$ queue. In \cite{multi-server_MMc}, the study was extended to the general M/M/$c$ queueing model where the newly arrived update can preempt the oldest update in service when all servers are busy. Considering the improvement of the robustness of the parallel transmission strategy to the system, some works studied the multi-queue parallel transmission system\cite{multi-queue_block-error,multi-queue_dual-queue_energy,multi-queue_DDP, multi-queue_hete-server,multi-queue_bound,multi-queue_multi-source}.
The dual-queue systems were studied in\cite{multi-queue_block-error,multi-queue_dual-queue_energy,multi-queue_DDP,multi-queue_hete-server}. The results showed that the average AoI performance of the systems in \cite{multi-queue_block-error,multi-queue_dual-queue_energy,multi-queue_DDP,multi-queue_hete-server} was significantly improved compared with corresponding single queue systems. The authors of \cite{multi-queue_bound} derived the statistical AoI bounds for a broad class of parallel systems.

In practice, the service capability of a single device, e.g., bandwidth, transmit power, is usually fixed or limited.
Then, to achieve a good performance regarding timeliness, using multiple devices for transmission is a viable solution.
It can be observed from the mentioned existing works that the multi-queue parallel transmission system (each queue uses independent channels to transmit updates) can not only improve the timeliness of information but also enhance the robustness of the system.
Motivated by this, we consider a status update system where multiple devices observe the same physical process and send updates to a common monitor.
In our previous works \cite{early_work-cl} and \cite{early_work-conf}, we derived the average AoI of a dual queue system and a general multi-queue system where all devices are the same type.
In this work, we extend the study to a multi-type-device network where multiple sources are sensed.
Specially, we consider two systems.
One uses only one type of device, while the other uses two types of devices.
The contributions of this work are summarized below:
\begin{itemize}
  \item We analyze the age performance of the \emph{Basic System} where only multiple type-1 devices sense a common physical process. The type-1 devices take the zero-wait (ZW) sampling policy and follow an exponentially distributed service time. Based on the SHS framework, the closed-form expression of the average AoI was derived. The theoretical results reveal that using multiple additional devices with the same service rate for parallel transmission can significantly improve timeliness compared with the single queue system.
  \item We investigate the age performance of the \emph{Hybrid System}, which incorporates multiple type-2 devices that transmit Poisson arrival updates (the Poisson process generates the updates) and have a negligible service time into the \emph{Basic System}. By leveraging the SHS framework, we derive the analytical expressions of the average AoI of the \emph{Hybrid System}. We show that the logarithm of average AoI is linear with the logarithm of the total arrival rate of the type-2 devices when only type-2 devices exist. In addition, it is found that the benefit brought by increasing unit device is higher for type-2 device compared to type-1 device.
  \item Numerical results confirm the accuracy of the average AoI expression. It is shown that the logarithm of average AoI is linear with the logarithm of the number of type-1 devices/total arrival rate of type-2 devices when the number of one of the two types of devices is fixed. It is also shown that there is a threshold interval for the budget for deployments to determine whether the optimal scheme is to deploy only one type of device or a hybrid deployment. Besides, the proposed system significantly outperforms the FCFS M/M/$N$ system in terms of average AoI. In addition, it achieves almost the same average AoI as that of the LCFS M/M/$N$ system with preemption under specific conditions.
\end{itemize}

\section{System Model}\label{sec_2}
In this section, we introduce the multi-queue status update system and the AoI metric.
\subsection{Multi-Queue Status Update System}
We consider a scenario where several external processes need to be sensed and sent to a common destination in a timely manner.
Specifically, we consider two different types of devices with different capabilities.
The type-1 devices are ad hoc devices located close to the specific external processes that only observe them; thus, they can generate updates and transmit through an erasure wireless channel to the destination.
On the other hand, type-2 devices are infrastructure-based devices located in fixed points.
In particular, the type-2 devices may be the sensors equipped at the monitor or remote data collector connected with the monitor through wired and dedicated cables.
The updates on type-2 devices are according to a Poisson process generated externally and forwarded to them.
Thus, type-2 devices can observe multiple processes.
This system showcases the interplay of the co-existence of different devices, and it can be applied to many scenarios and provide insights.
Consider the following two examples as potential application scenarios for the modeled system.
The first application example is the intelligent agricultural system, which requires the detection of various environmental parameters (such as light intensity, temperature, and humidity).
To provide comprehensive and timely guidance, multiple devices for each parameter must be set up, and setting up common equipment that enables the observation of multiple parameters can further improve their timeliness.
The second application example is the building fire prevention system.
In this system, deploying multiple temperature monitoring devices on each floor and transmitting devices connected to multiple floors can effectively prevent emergencies and ensure the safety of people and their property.

For a specific source (an external process), a multi-queue status update system can be utilized, as shown in Fig.~\ref{SystemModel}.
There are $N$ type-1 devices and $M$ type-2 devices ($N\in \mathbb{N}^{*},M\in\mathbb{N}$).
Denote the index set of the $N$ type-1 devices and $M$ type-2 devices by $\mathcal{N} = \{1,~2,~\cdots,~N\}$ and $\mathcal{M} = \{0,~1,~\cdots,~M\}$ respectively.
The $N$ type-1 devices can generate an update at will since they are ad hoc devices placed near the source.
In particular, taking the complexity and the performance of the policy into account, we assume that the type-1 devices employ the ZW policy for sampling.
The ZW policy has low computational complexity.
It is applied in IoT scenarios and can achieve an information freshness performance that is acceptable\cite{queue_schedule_ZW}.
Under the ZW policy, each type-1 device samples a new update and sends it to the monitor when the service of the previous update is just completed.
Taking channel fading and path loss in wireless networks into account, the service time of type-1 device $i$ $(i\in\mathcal{N})$ is assumed to follow the exponential distribution with parameter $\mu_{i}$.
Denote the total service rate of type-1 devices by $\mu$, i.e., $\sum_{i=1}^{N}\mu_{i}=\mu$.
Type-2 devices' service time, i.e., transmission time is much shorter than that of type-1 devices and is assumed to be negligible based on its properties.
On the other hand, the generation of an update of the type-2 device $i$ $(i\in\mathcal{M})$ is assumed to be a Poisson process with parameter $\lambda_{i}$ since the type-2 devices are infrastructure-based devices located in fixed points.
Denote the total arrival rate of type-2 devices by $\lambda$, i.e., $\sum_{i=1}^{M}\lambda_{i}=\lambda$.

Note that for a source, no additional type-2 devices are likely to sense it, i.e., $M=0$. Accordingly, we will analyze the average AoI of the considered system separately according to the existence of type-2 devices.
For clarity, we represent the considered system in two cases
\begin{itemize}
  \item \emph{Basic System}: Only $N$ type-1 devices are equipped, i.e., $M=0$.
  \item \emph{Hybrid System}: Besides of $N$ type-1 devices, there are additional type-2 devices, i.e., $M>0$.
\end{itemize}

\subsection{AoI Metric}
We take the average AoI as the performance metric to measure the timeliness of the considered system. The instantaneous age of the monitor at time $t$ is defined as $\Delta(t)=t-U(t)$ where $U(t)$ is the time stamp of the latest update adopted by the monitor rather than received which is different from the single queue system. Due to the randomness of the service time, some updates may be outdated when they arrive at the monitor. Since only a fresh update is useful for monitor, the monitor will compare its AoI with the AoI of the newly arrived update. The newly arrived update will be adopted only if its AoI is smaller than that of the monitor. Then the average AoI of the considered system can be expressed as
\begin{equation}\label{average age}
 {\Delta}_{}=\lim_{\tau\rightarrow\infty}\frac{1}{\tau}\int_{0}^{\tau}\Delta(t)\,\mathrm{d}t.
\end{equation}

\section{SHS Analysis Framework}\label{sec_3}
Considering the difficulty of age analysis caused by out of order arrival in multi-queue system and the effectiveness of SHS method in simplifying AoI analysis, we study the AoI of the considered system under SHS framework.
The SHS method models a stochastic system by dividing the state into two components $q(t)$ and $\boldsymbol{x}(t)$, where $q(t)\in\mathcal{Q}=\{1,~2,~\cdots,~m\}$ is a continuous-time finite-state Markov chain that describes the evolution of the system and $\boldsymbol{x}(t)=[x_{0}(t)\cdots x_{n}(t)]\in\mathbb{R}^{n+1}$ is a continuous process that describes the evolution of a set of age-related processes.

The Markov chain $q(t)$ is determined by the transition set $\mathcal{L}$.
Specifically, a transition occurs when an update just arrives or completes the service.
For a transition $l\in\mathcal{L}$, the states before and after the transition are denoted by $q_{l}$ and $q'_{l}$, respectively, and its occurrence rate is denoted by $\lambda^{(l)}$.
When transition $l$ occurs, the discrete state $q_{l}$ jumps to $q'_{l}$ and the continuous state is reset from $\boldsymbol{x}$ to $\boldsymbol{x}'=\boldsymbol{x}\boldsymbol{A}_{l}$, where $\boldsymbol{A}_{l}\in\mathbb{R}^{(n+1)\times(n+1)}$ is the reset map of transition $l$.
It is worth noting that for a transition $l$, $q_{l}$ and $q'_{l}$ can be the same, which is called self-transition.
In this case, a reset occurs in the continuous state while the discrete state remains the same.

In a specific state $q\in\mathcal{Q}$, the continuous process evolves following the differential equation $\dot{\boldsymbol{x}}=\boldsymbol{b}_{q}=[b_{q0}\quad b_{q1} \cdots b_{qn}]$.
As the continuous process describes the evolution of age-related processes, $b_{qi}$ $(i\in \{0\}\cup \mathcal{N})$ would be binary.
Specifically, if an age process $x_{i}$ increases at a unit rate in state $q$, then $b_{qi}=1$.
If $x_{i}$ keeps invariant, then $b_{qi}=0$.
For clarity, the incoming and outgoing transitions sets are defined for each state $q$ as $\mathcal{L}'_{q}=\{l\in \mathcal{L}:q'_{l}=q\}$, $\mathcal{L}_{q}=\{l\in \mathcal{L}:q_{l}=q\}$, respectively.

The SHS method introduces the correlation vector between the age  process $\boldsymbol{x}(t)$ and the discrete state $q$ to compute the average AoI.
Let $\pi_{q}$ represent the probability of the Markov chain being in state $q$ and $\boldsymbol{v}_{q}$ the correlation vector. Then, $\pi_{q}$ and $\boldsymbol{v}_{q}$ are defined as
\begin{align}
\label{state probabilities}\pi_{q}(t)&=\mathbb{E}[\delta_{q,q(t)}]=\mathbf{P}{\mathbf{r}}[q(t)=q],\\
\label{correlation}\boldsymbol{v}_{q}(t)&=[v_{q0}(t) \quad v_{q1}(t) \quad ... \quad v_{qn}(t)]=\mathbb{E}[\boldsymbol{x}(t)\delta_{q,q(t)}].
\end{align}

The steady state probability vector $\overline{\boldsymbol{{\pi}}} = [\overline{\pi}_{0} ~ \cdots ~ \overline{\pi}_{m}]$ exists\footnote{Under ergodicity assumptions as in \cite{multi-source_SHS-propose}.} and is the solution of the following equations
\begin{subequations}
\begin{align}
\overline{\pi}_q\sum_{l\in \mathcal{L}_{q}}\lambda^{(l)}&=\sum_{l\in \mathcal{L}'_{q}}\lambda^{(l)}\overline{\pi}_{q_l},\quad q\in\mathcal{Q},\label{steady state probability1}\\
\sum_{q\in\mathcal{Q}}\overline{\pi}_{q}&=1. \label{steady state probability2}
\end{align}
\end{subequations}
Accordingly, the correlation vector $\boldsymbol{v}_{q}(t)$ converges to a non-negative limit $\overline{\boldsymbol{v}}_{q}$ satisfying\cite[Theorem 4]{multi-source_SHS-propose}
\begin{equation}\label{}
  \lim_{t\rightarrow\infty}\boldsymbol{v}_{q}(t)=\overline{\boldsymbol{v}}_{q}, \quad q \in \mathcal{Q}.
\end{equation}
\begin{equation}\label{correlation vector}
\overline{\boldsymbol{v}}_q\sum_{l\in \mathcal{L}_q}\lambda^{(l)}=\boldsymbol{b}_{q}\overline{\pi}_{q}+\sum_{l\in\mathcal{L}'_{q}}\lambda^{(l)}{\overline{\boldsymbol{v}}}_{q_l}\boldsymbol{A}_{l},\quad q\in\mathcal{Q}.
\end{equation}
Since $\mathbb{E}[\boldsymbol{x}(t)]=\sum_{q\in\mathcal{Q}}\mathbb{E}[\boldsymbol{x}(t)\delta_{q,q(t)}]=\sum_{q\in\mathcal{Q}}\boldsymbol{v}_{q}(t)$, it has
\begin{equation}\label{E[x]}
  \mathbb{E}[\boldsymbol{x}]=\lim_{t\rightarrow\infty}\mathbb{E}[\boldsymbol{x}(t)]=\lim_{t\rightarrow\infty}\sum_{q\in\mathcal{Q}}\boldsymbol{v}_{q}(t)
                            =\sum_{q\in\mathcal{Q}}\overline{\boldsymbol{v}}_{q}.
\end{equation}
Then, the average AoI of the system can be expressed as
\begin{equation}\label{average AoI}
\Delta=\mathbb{E}[x_{0}]=\sum_{q\in\mathcal{Q}}\overline{v}_{q0}.
\end{equation}
As \eqref{average AoI} implies, to get the average AoI of the system, it is equivalent to derive $\overline{v}_{q0}$ for all states.

\section{Average AoI of the Multi-Queue Status Update System}
In this section, based on the SHS framework and following its required steps, we derive the closed-form expressions of the average AoI for the \emph{Basic System} and the \emph{Hybrid System}.
In addition, for the two systems, we further derive the simplified forms of the average AoI under homogeneous condition.

\subsection{The Average AoI of the Basic System}\label{subsec:Sd}
Under the SHS framework, the first step is to select the appropriate discrete state set and age-related processes.
Then, according to the discrete states and the selected age-related processes, all SHS elements are defined, including all transitions, their corresponding parameters (transition rate and reset map), and the differential equations describing the evolution of the age-related processes.
After that, it is required to find the steady state probability vector of the Markov chain for the discrete states.
Finally, by solving the correlation vector with \eqref{correlation vector}, one can get the average AoI of the system based on \eqref{average AoI}.

Let us start with selecting the appropriate discrete state set $\mathcal{Q}$ and age-related processes $\boldsymbol{x}(t)$.
Note that the AoI at the monitor is equal to the AoI of the update which has the smallest AoI among all updates received before a new update arrives.
Since the update with the smallest AoI can come from any one of the $N$ type-1 devices, we define the discrete state as the source index tuple of the latest $N$ updates.
Formally, we define the discrete state as a vector $\boldsymbol{q}=[q(1)~\cdots ~ q(N)]\in\mathcal{N}^{N}$ where $q(i)$, $i\in\mathcal{N}$ records the index of the type-1 device the $i$-th latest update came from, regardless of the adoption of update or not at the monitor.
\begin{figure}[t]
\centering
\subfigure[Set-A: Without self-transition ] 
{\includegraphics[width=115pt]{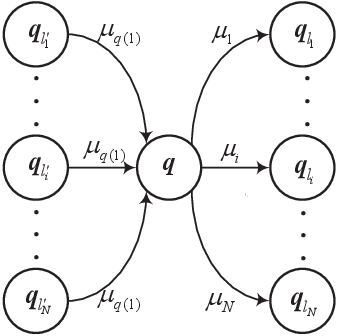}} 
\subfigure[Set-B: With self-transition  ]  
{\includegraphics[width=115pt]{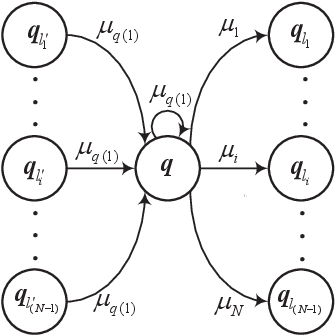}}
\caption{Local Markov chain transition of the \emph{Basic system}.} 
\label{transitions of q}  
\vspace{-1em}
\end{figure}

Note that there are $N^{N}$ discrete states in total.
We divide all the discrete states into two sets and introduce the state transitions for state in each set with the help of the local Markov chain transition diagram as shown in Fig.~\ref{transitions of q}.
Specifically, according to the occurrence of self-transition at a state, we consider set-A states as those without self-transition and set-B states as those with self-transition.
In fact, both sets of states have $N$ incoming transitions whose transition rates are all $\mu_{q(1)}$ and $N$ outgoing transitions whose transition rates are $\mu_{1}$ to $\mu_{N}$ respectively.
The difference between the two sets of states is that one of the incoming transitions and one of the outgoing transitions of the set-B states merge and form a self-transition with a rate of $\mu_{q(1)}$.

In particular, to judge whether the AoI of the monitor would be refreshed or not when an update is delivered, we introduce an auxiliary function $p(\boldsymbol{q})$ for state $\boldsymbol{q}$ to indicate the position where the source of the update repeats in the precedent updates for the first time.
Formally, $p(\boldsymbol{q})$ is expressed as
\begin{equation}\label{p(q)}
p({\boldsymbol{q}}) = 1+ \max_{r\in\{\bar{r}\in\mathcal{N}:\forall j,h \in \{ 1, \cdots ,\bar{r}\} , ~q(j) \ne q(h) ~ \text{if} ~ j \ne h\}}r
\end{equation}
One can see that $p(\boldsymbol{q})\in \{2,~3,~\cdots,~N+1\}$.
The continuous process is considered as $\boldsymbol{x}(t) =[x_{0}(t) \quad x_{1}(t)\quad \cdots \quad x_{N}(t)]$ where $x_{0}(t)$ records the age of the monitor at time $t$ and $x_{i}(t)$ $(i\in \mathcal{N})$ the age of the update generated by type-1 device $i$ at time $t$.

Based on the above definition, the average AoI of the \emph{Basic System} is given by the following theorem.
\theoremstyle{plain}
\newtheorem{mythm}{Theorem}
\begin{mythm}
The average AoI of the Basic $N$-queue $(N\in\mathbb{N}^{*})$ status update system is
\begin{equation}\label{nstreamaverageAoI}
  {\Delta}_{N}^{\text{Basic}}=\sum_{\boldsymbol{q}\in\mathcal{Q}}\frac{p(\boldsymbol{q})M(\boldsymbol{q})}{\mu^{N+1}},
\end{equation}
where $\mu = \sum_{i=1}^{N}\mu_{i}$ and $M(\boldsymbol{q})=\prod_{i=1}^{N}\mu_{q(i)}$.
\end{mythm}

\begin{IEEEproof}
See Appendix \ref{Theorem_1_proof}.
\end{IEEEproof}
\newtheorem{corollary}{Corollary}
\begin{corollary}
In the homogeneous Basic $N$-queue status update system where $\mu_{i}={\mu}/{N}$~$(i\in \mathcal{N})$, the average AoI is
\begin{equation}\label{corollary_1}
   \Delta _{N}^{\text{Bas-Homo}} = \frac{N}{\mu}\sum\limits_{r = 1}^{N } \frac{\binom{N-1}{r-1} (r+1)!}{N^{r+1}}.
\end{equation}

\end{corollary}
\begin{IEEEproof}
See Appendix \ref{Corollary_1_proof}.
\end{IEEEproof}
Based on Corollary 1, we present the ratio of the average AoI of the homogeneous \emph{Basic System} to that of the single queue system under some typical values of $N$ in Tab.~\ref{table_Corollary_1}.

\begin{table}[t]
\centering
\caption{The Ratio of The Average AoI of The Homogeneous Basic System to That of The Single Queue System, i.e., $\Delta _{N}^{\text{Bas-Homo}}/\Delta _{1}^{\text{Bas-Homo}}$.}
\label{table_Corollary_1}
\begin{tabular}{|c|c|c|c|}
  \hline
  $N$ & Ratio & $N$ & Ratio \\
  \hline
  1 & 100\% & 10 & 23.30\% \\
  \hline
  2 & 62.5\% & 100 & 6.61\% \\
  \hline
  3 & 48.15\% & 1000 & 2.02\% \\
  \hline
  4 & 40.23\% & 10000 & 0.63\% \\
  \hline
\end{tabular}
\end{table}

\subsection{The Average AoI of the Hybrid System}
In the \emph{Hybrid System}, the joining of the type-2 device not only increases the state space, which brings potential complexity improvement of analysis for SHS elements, but also proposes a new challenge to analysis of the first component of the correlation vector $\overline{v}_{\boldsymbol{q}0}$.

Since the service time of the type-2 device is negligible, the update from the type-2 device will certainly be adopted by the monitor, and the monitor's age will be reset to zero after receiving the update from the type-2 device.
Thus, updates from different type-2 devices have the same effect on the monitor's AoI, i.e., they are equivalent for the system.
As the sum process of Poisson processes is still Poisson, one can regard the $M$ type-2 devices as one type-2 device with arrival rate $\lambda$.
Without loss of generality, we consider $M=1$ in the following analysis.

Similar to the \emph{Basic System} case, to derive the average AoI based on the SHS framework, it first needs to select appropriate discrete state set $\mathcal{Q}$ and age-related processes $\boldsymbol{x}(t)$.
Since the monitor would always adopt the update from the type-2 device, we still consider the discrete state as the source index tuple of the latest $N$ updates. Formally, the discrete state is defined as a vector $\boldsymbol{q}=[q(1) ~ q(2) ~ \cdots ~ q(N)]\in(\mathcal{N}\cup \{N+1\})^{N}$ where $q(i)$ $(i\in\mathcal{N})$ records the index of the device the $i$-th latest update came from\footnote{For clarity, we call the $N$ type-1 devices as device $1$ to device $N$, and the only type-2 device as device $(N + 1)$.}.
In addition, to uniformly describe the service rate of the $N$ type-1 devices and the arrival rate of the type-2 device, we define a variable $u_{i}~(i\in\mathcal{N}\cup \{N+1\})$ as

\begin{equation}\label{}
  u_{i}=
  \begin{cases}
  \mu_{i},&i\in\mathcal{N},\\
  \lambda,&i=N+1.
  \end{cases}
\end{equation}

\begin{figure}[t]
\centering
\subfigure[Class-A: Without self-transition ] 
{\includegraphics[width=115pt]{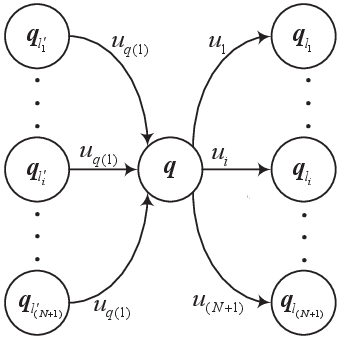}} 
\subfigure[Class-B: With self-transition  ]  
{\includegraphics[width=115pt]{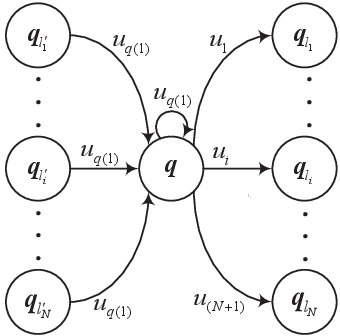}}
\caption{Local Markov chain transition of the \emph{Hybrid system}.} 
\label{n_1 stream transition}  
\vspace{-1em}
\end{figure}
By definition, there are $(N+1)^{N}$ discrete states in total.
As the local Markov chain shown in Fig.~\ref{n_1 stream transition}, we divide all the discrete states into two classes, namely, class-A and class-B, according to the occurrence of self-transition at a state.
Both classes of states have $N+1$ incoming transitions whose transition rates are all $u_{q(1)}$ and $N+1$ outgoing transitions whose transition rates are $u_{1}$ to $u_{(N+1)}$ respectively, while one of the incoming transitions and
one of the outgoing transitions of the class-B states merge and form a self-transition with a rate of just $u_{q(1)}$.

The continuous processes are considered as $\boldsymbol{x}(t) =[x_{0}(t)$ $ ~ x_{1}(t)~ \cdots ~ x_{N}(t)~x_{N+1}(t)]$ where $x_{0}(t)$ is the age of the monitor at time $t$ and $x_{i}(t)$ $(i\in \mathcal{N}\cup\{N+1\})$ is the age of the update generated/adopted by device $i$ at time $t$.
Since the service time of device $(N+1)$ (the only type-2 device) is negligible, we know that $x_{N+1}(t)$ is always zero.
Thus, we ignore the analysis of $x_{N+1}(t)$ in the following, i.e., we just consider $\boldsymbol{x}(t) =[x_{0}(t) ~ x_{1}(t)~ \cdots ~ x_{N}(t)]$.

Similar to the \emph{Basic System}, two auxiliary functions $p(\boldsymbol{q})$ and $\tilde{p}(\boldsymbol{q})$ for status $\boldsymbol{q}$ are defined to help judge whether the AoI of the monitor will be refreshed or not.
The function $p(\boldsymbol{q})$ which indicates the position where the source of the update first repeats in the previous updates is the same as defined in \eqref{p(q)}\footnote{We force to have that $p(\boldsymbol{q})=N+1$ when the sources of the previous $N$ updates are different such that the definition of $p(\boldsymbol{q})$ in the Hybrid system is the same as that in the Basic system.}.
The function $\tilde{p}(\boldsymbol{q})$ which indicates the position of the latest update that came from device $(N+1)$ is defined as\footnote{When there is no update arrived from device $(N+1)$ in previous $N$ updates, the position of the latest update that came from device $(N+1)$ cannot be known, and we consider $\tilde{p}(\boldsymbol{q})=N+1$ in this case.}
\begin{equation}\label{}
  \tilde{p}(\boldsymbol{q})=
  \begin{cases}
  \min\limits_{r\in\{\overline{r}\in\mathcal{N}:q(\overline{r})=N+1\}}r,&\exists j\in\mathcal{N},q(j)=N+1,\\
  N+1,& \text{otherwise}.
  \end{cases}
\end{equation}

Based on the definition of the discrete state and the continuous processes, the average AoI of the \emph{Hybrid System} is given by the following theorem.
\begin{mythm}
The average AoI of the Hybrid $(N,M)$-queue $(N,M\in\mathbb{N}^{*})$ status update system is
\begin{equation}\label{n_1streamaverageAoI}
  \Delta_{N,M}^{\text{Hybrid}}=\sum_{\boldsymbol{q}\in\mathcal{Q}}\frac{\min\{p(\boldsymbol{q}),\tilde{p}(\boldsymbol{q})\}U(\boldsymbol{q})}{(\mu+\lambda)^{(N+1)}},
\end{equation}
where $\mu = \sum_{i=1}^{N}\mu_{i}$, $\lambda = \sum_{i=1}^{M}\lambda_{i}$ and $U(\boldsymbol{q})=\prod_{i=1}^{N}u_{q(i)}$.
\end{mythm}

\begin{IEEEproof}
See Appendix \ref{Theorem_2_proof}.
\end{IEEEproof}

\begin{corollary}
In the homogeneous Hybrid System where the service rate of all type-1 devices is the same, i.e. $\mu_{i}=\mu/N~(i\in\mathcal{N})$, the average AoI is
\begin{align}\label{collary2}
  \Delta_{N,M}^{\text{Hyb-Homo}} ={}&  \frac{N}{\mu}\sum\limits_{r = 1}^{N } \frac{\binom{N-1}{r-1} (r+1)!}{N^{r+1}}\left(\frac{\mu}{\mu+\lambda}\right)^{N+1}\notag\\
   &  + \sum_{k=1}^{N}(\mathbb{I}_{\{N-k-1>0\}}A(k)+B(k)),
\end{align}
where $\mathbb{I}_{\{N-k-1>0\}}$ is equal to 1 when $N-k-1 > 0$ and 0 otherwise,
\begin{align}\label{}
  A(k) & =\sum_{p=2}^{N-k}\frac{\binom{N-p}{k}\binom{N-1}{p-2}p!\mu^{N-k}\lambda^{k}}{N^{p-1}(\mu+\lambda)^{N+1}}, \notag\\
  B(k) & =\sum_{p=1}^{N-k+1}\frac{\binom{N-p}{k-1}\binom{N}{p-1}p!\mu^{N-k}\lambda^{k}}{N^{p-1}(\mu+\lambda)^{N+1}}.
\end{align}
\end{corollary}

\begin{IEEEproof}
See Appendix \ref{Corollary_2_proof}.
\end{IEEEproof}

\newtheorem{remark}{Remark}
Based on Corollary 2, we present the ratio of the average AoI  of the homogeneous \emph{Hybrid System} to that of the \emph{Hybrid System} with $N=1$ and $M=1$ under different $N$ and $M$ in Tab.~\ref{table_Corollary_2}.
Note that $\lambda_{i}=1$ is adopted in Tab.~\ref{table_Corollary_2}.
Accordingly, it has $\lambda=M\lambda_{i}=M$.
\begin{table}[t]
\centering
\caption{The Ratio of The Average AoI of the Homogeneous Hybrid System to That of The Hybrid System with $N=1$ and $M=1$, i.e., $\Delta_{N,M}^{\text{Hyb-Homo}}/\Delta_{1,1}^{\text{Hyb-Homo}}$.}
\label{table_Corollary_2}
\begin{tabular}{|c|c|c|c|c|c|}
\hline
\diagbox{$N$}{$M$} & 1 & 5 & 10 & 20 & 30\\
\hline
1 & 100\% & 25.93\% & 13.22\% & 6.65\% & 4.43\% \\
\hline
10 & 46.70\% & 21.66\% & 12.39\% & 6.52\% & 4.40\% \\
\hline
100 & 16.20\% & 12.05\% & 8.91\% & 5.66\% & 4.08\% \\
\hline
500 & 7.38\% & 6.42\% & 5.49\% & 4.18\% & 3.32\% \\
\hline
1000 & 5.24\% & 4.75\% & 4.23\% & 3.43\% & 2.86\% \\
\hline
\end{tabular}
\end{table}

\begin{remark}
Let the service rate of all type-1 devices in Hybrid System be zero, it is equivalent to the system with only type-2 devices.
In this case, the average AoI is $1/\lambda$.
\end{remark}

\section{Numerical Results}\label{sec_4}
In this section, we provide numerical results to validate the accuracy of the analysis and evaluate the information freshness of the considered systems.

\begin{figure}[t]
\centering
\subfigure[$\lambda = 0$ ] 
{\label{sim1}\includegraphics[width=0.45\textwidth]{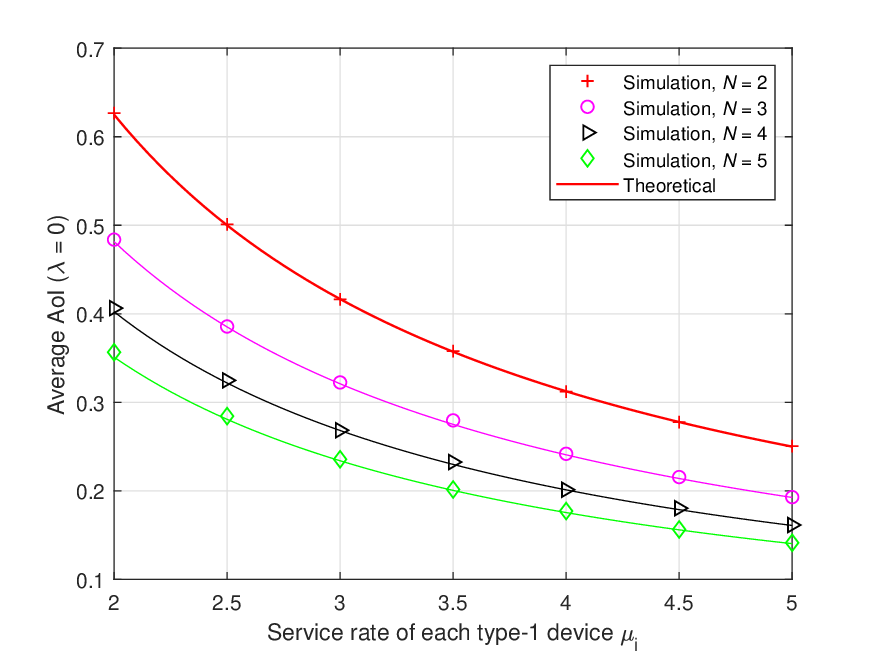}} 
\subfigure[$\lambda = 1$]  
{\label{sim2}\includegraphics[width=0.45\textwidth]{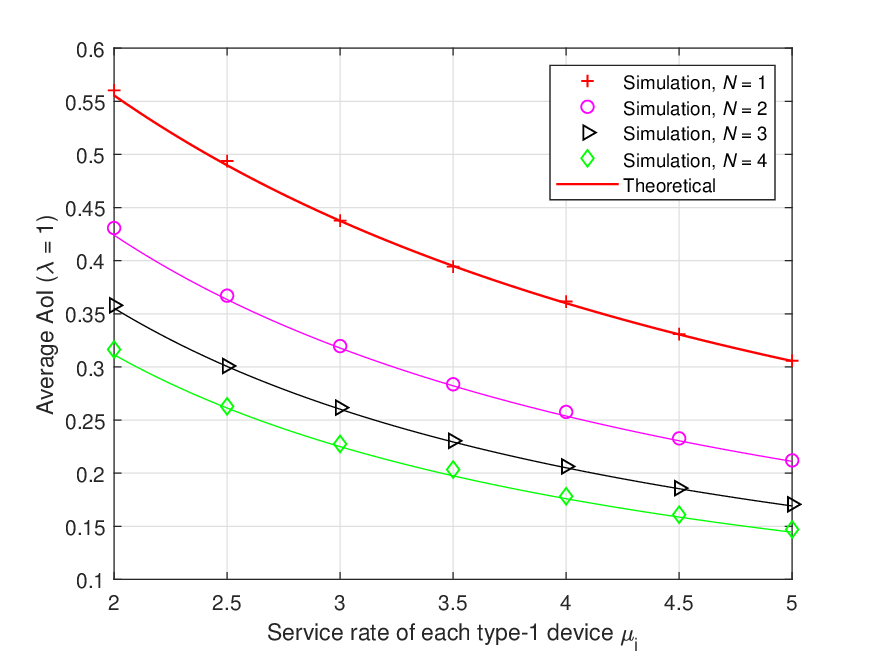}}
\caption{Average AoI of the multi-queue status update system.} 
\label{G_sim}  
\vspace{-1em}
\end{figure}
For $\lambda = 0$ and $\lambda = 1$, we depict average AoI $\Delta_{N,M}^{\text{Hyb-Homo}}$ (when $\lambda = 0$, $\Delta_{N,M}^{\text{Hyb-Homo}}$ degenerates to $\Delta_{N}^{\text{Bas-Homo}}$) versus the service rate of single type-1 device $\mu_{i}$ for different $N$ based on \eqref{collary2} in Fig.~\ref{sim1} and Fig.~\ref{sim2} respectively.
The corresponding simulation results are shown as markers.
It can be seen that the theoretical results coincide with the simulation results, which verifies the reliability of our derivation.
It can be also observed from Fig.~\ref{G_sim} that the maximum of AoI reduction is brought by the first additional type-1 device and the benefit obtained from each additional type-1 device is gradually decreasing.

\begin{figure}[t]
\centering
\subfigure[$\lambda = 0$ ] 
{\label{vsN_1}\includegraphics[width=0.45\textwidth]{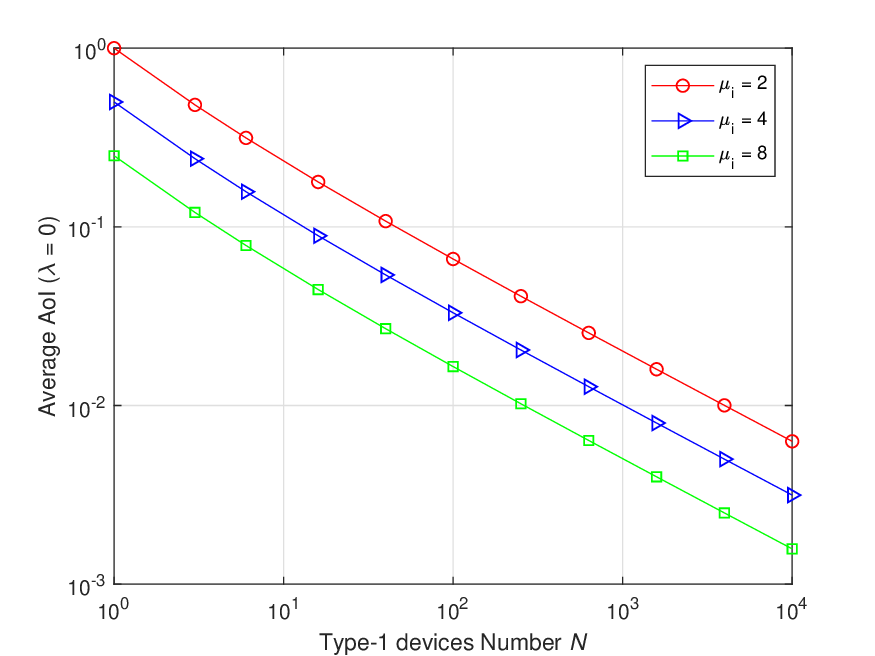}} 
\subfigure[$\lambda > 0$]  
{\label{vsN_3}\includegraphics[width=0.45\textwidth]{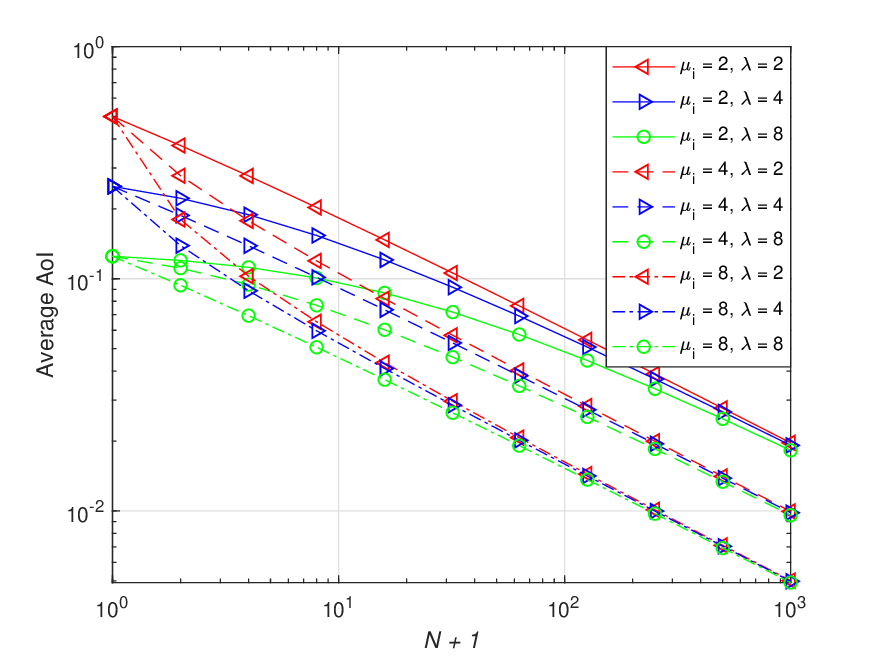}}
\caption{The average AoI vs. the type-1 device number.} 
\label{AoIvsN}  
\vspace{-1em}
\end{figure}

Fig.~\ref{AoIvsN} shows how the average AoI $\Delta_{N,M}^{\text{Hyb-Homo}}$ varies with the number of type-1 devices $N$.
It is noteworthy that the logarithm of average AoI is approximately linear with the logarithm of the number of type-1 device $N$ for $\lambda = 0$.
An increase of 100 times in the number of type-1 device brings 10 times reduction in the average AoI.
For $\lambda > 0$, from Fig.~\ref{vsN_3}, it can be seen that the above phenomenon still holds when $N$ is large.
In addition, one can find that for the same $\mu_{i}$, the logarithm of the average AoI of the system with different $\lambda$ tends to the same straight line when the $N$ is large.
This is because when $N$ is large, the updates from type-2 devices account for a small proportion of all updates received by the monitor, i.e., the effect of the type-2 devices is small for the system.

\begin{figure}[t]
  \centering
  \includegraphics[width=0.46\textwidth]{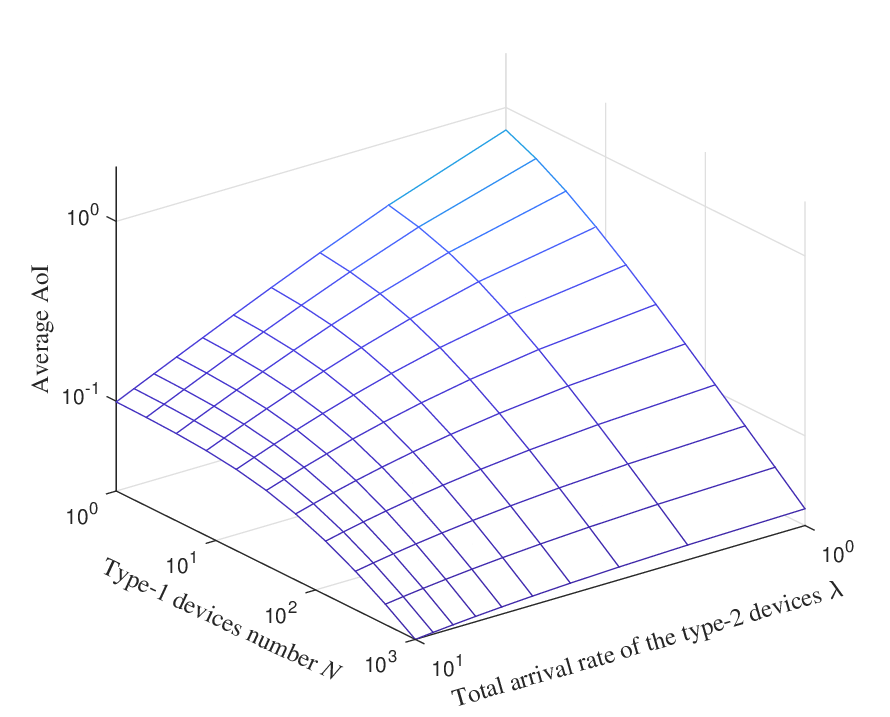}\\
  \caption{The average AoI vs. the total arrival rate of the type-2 devices.}\label{AoIvsNM}
  \vspace{-1em}
\end{figure}

Fig.~\ref{AoIvsNM} shows the relationship between the average AoI and the total arrival rate of the type-2 devices $\lambda$ (the service rate of single type-1 device is set as $\mu_{i} = 1$).
It can be found that when the number of type-1 devices is fixed, the logarithm of average AoI and the logarithm of the total arrival rate of the type-2 devices are linear when $\lambda$ is relatively large.
In addition, it can be observed that the benefit brought by increasing unit device is higher for type-2 device compared to type-1 device, i.e., more type-1 devices are required to achieve the same AoI performance.
When the total service rate of type-1 devices $\mu$ is less than $\lambda$, the gain of reduced the average AoI obtained from additional type-1 devices disappears.
While when $\lambda$ is less than $\mu$, an increase in $\lambda$ can still bring some gain.

\begin{figure}[t]
  \centering
  \includegraphics[width=0.46\textwidth]{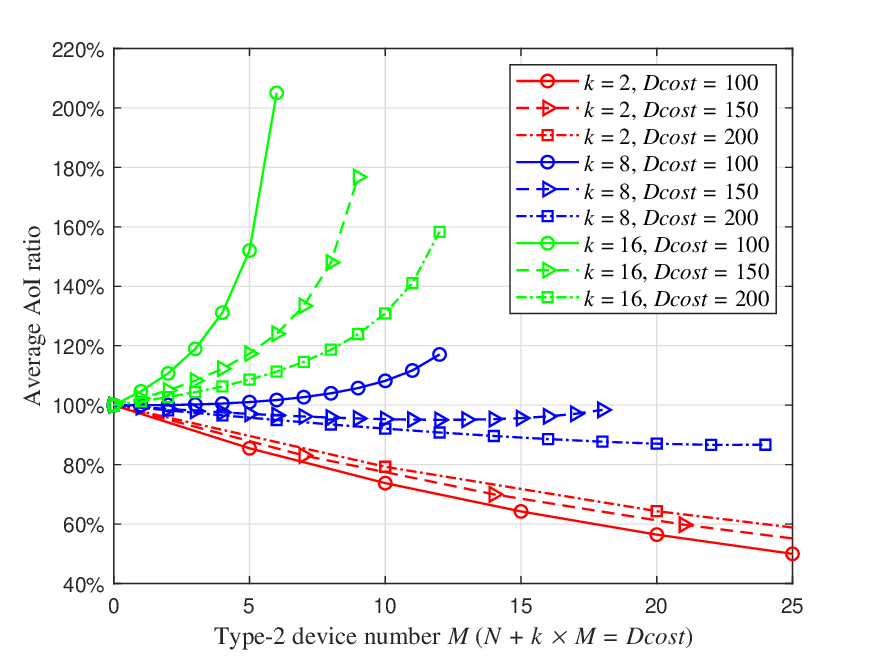}\\
  \caption{The average AoI under different deployment schemes.}\label{AoIvsDcost}
  \vspace{-1em}
\end{figure}

\begin{figure}[t]
  \centering
  \includegraphics[width=0.46\textwidth]{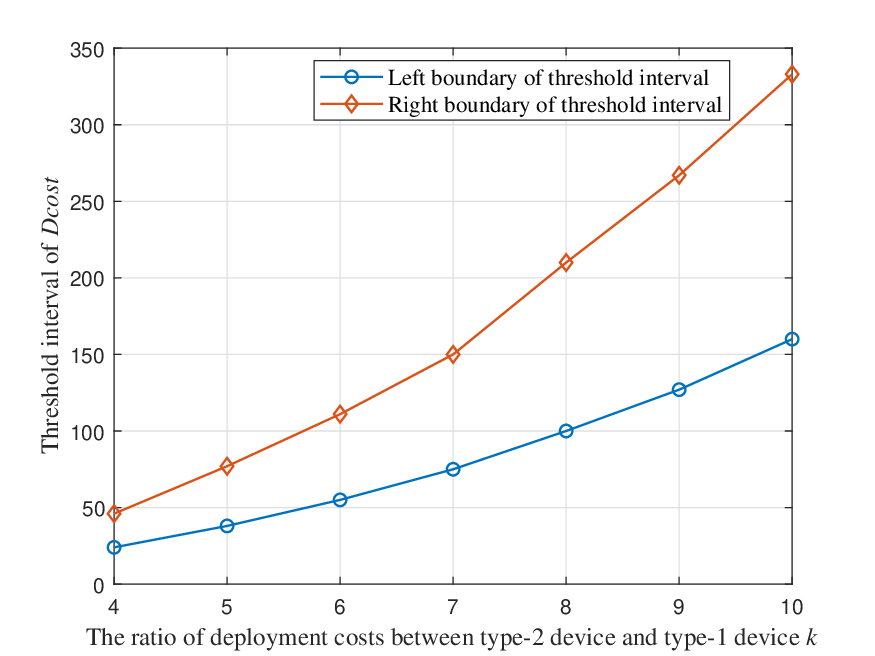}\\
  \caption{The threshold interval of $Dcost$ vs $k$.}\label{kvsThreshold}
  \vspace{-1em}
\end{figure}

Considering that the budget for deployment is usually limited in practice, we depict the average AoI performance under different deployment schemes for several cost of deployment units in Fig.~\ref{AoIvsDcost}\footnote{To demonstrate the average AoI performance under differences deployment schemes, the ordinate is considered as the ratio of the average AoI achieved by the specific deployment scheme to that achieved by the deployment scheme of $M = 0$, i.e., $\Delta_{N,M}^{\text{Hyb-Homo}}/\Delta_{Dcost,0}^{\text{Hyb-Homo}}$.} where $k$ represents multiple of the deployment cost of the type-2 device compared to the type-1 device (the deployment cost of the type-1 device has been normalized) and $Dcost$ represents the total cost of deployment units.
It can be found that when $k = 2~(16) $, the optimal deployment scheme is to deploy only type-2 (type-1) devices under three values of $Dcost$, i.e., 100, 150, and 200.
In contrast, when $k = 8$, the optimal deployment scheme will present different situations based on different values of $Dcost$.
Thus, there exists a threshold, or more precisely, a threshold interval for a fixed $k$.
When $Dcost$ is less (greater) than the minimum (maximum) value of the threshold interval, the optimal deployment scheme is to deploy only type-1 (type-2) devices.
When it is in the threshold interval, the optimal deployment scheme is composed of two types of device.
Then, under the same parameters, we further simulated the threshold interval of $Dcost$ corresponding to different $k$ values, as shown in Fig.~\ref{kvsThreshold}.
It can be found that the threshold interval of $Dcost$ increases with the increase of $k$.
This is because that as $k$ increases, the advantage in the number of type-1 devices under the same $Dcost$ will be greater, resulting in a higher threshold interval.
Another finding is that the right boundary of the threshold interval is approximately twice that of the left boundary for the considered $k$, $\mu_{i}$ and $\lambda_{i}$.

\begin{figure}[t]
  \centering
  \includegraphics[width=0.46\textwidth]{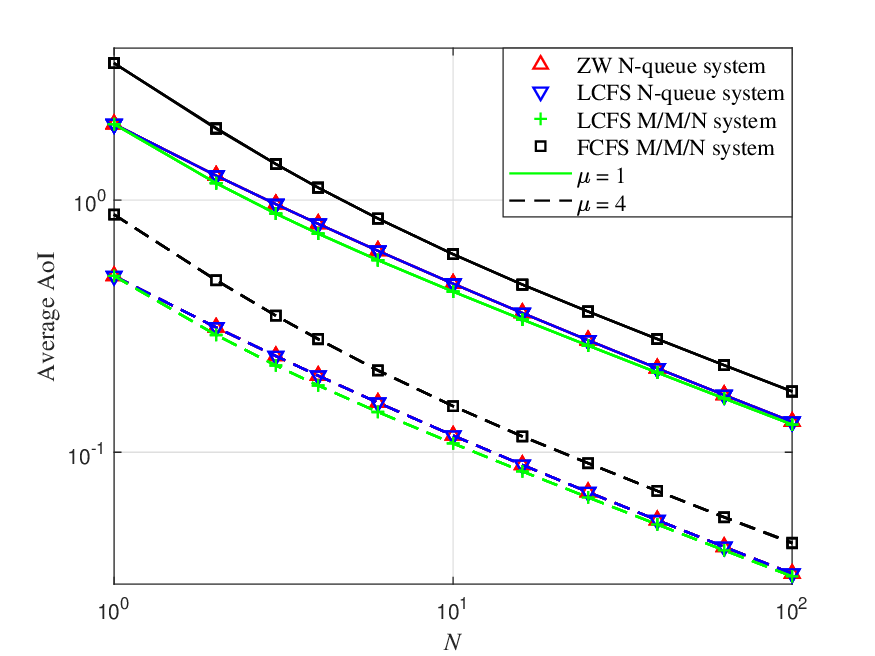}\\
  \caption{Comparison of the average AoI performance of the considered system and other systems.}
  \label{compare1}
  \vspace{-1em}
\end{figure}

Finally, we compare the average AoI performance of our system ($\lambda = 0$, refer to ZW $N$-queue system) with that of the multi-queue system where each queue is a LCFS queue with preemption (refer to LCFS $N$-queue system)\cite{multi-queue_multi-source}, the LCFS M/M/$N$ system with preemption (refer to LCFS M/M/$N$ system) \cite{multi-server_MMc} and the FCFS M/M/$N$ system (refer to FCFS M/M/$N$ system) in Fig.~\ref{compare1}.
In these four systems, two settings of service rates are considered, i.e., $\mu_{i} = \mu/N = 1$ and $\mu_{i}=4$.
The load of the LCFS $N$-queue system, the LCFS M/M/$N$ system, and the the FCFS M/M/$N$ system are set as $\rho=\frac{\lambda}{\mu_{i}}=1$, $\rho=\frac{\lambda}{N\mu_{i}}=1$, and $\rho=\frac{\lambda}{N\mu_{i}}=0.56$, respectively\footnote{In our system, it can be considered that each type-1 device has an arrival rate equal to the service rate, then the equivalent load is $\rho=\frac{\mu_{i}}{\mu_{i}}=1$.
Thus we set the load of the LCFS $N$-queue system and the LCFS M/M/$N$ system as 1.
The FCFS M/M/$N$ system had not been studied, thus we use the optimal load $\rho = 0.56$ for $N=2$, i.e., the FCFS M/M/$2$ system, instead\cite{multi-server_MM2_journal}.
Besides, the simulation results of the FCFS M/M/$N$ system rather than the theoretical one is used for comparison in Fig.~\ref{compare1}.}.
An interesting discovery is that all four systems have the property that the logarithm of the average AoI is approximately linear with the logarithm of the number of devices.
Our system, the LCFS $N$-queue system and the LCFS M/M/$N$ system have almost the same average AoI performance and are better than the FCFS M/M/$N$ system.
Note that the average AoI of the preemption system (the LCFS $N$-queue system and the LCFS M/M/$N$ system) is related to the load.
The preemption system can provide a greater arrival rate to achieve a larger load and thereby improve the performance.
However, it has a high cost to maintain excessive sampling and updating.
Meanwhile, the preemptive system requires extra computing resources on servers.
As the computing resources are limited in the IoTs, the ZW discipline is more suitable for status sampling.

\section{Conclusion}\label{sec_5}
This paper investigated the average AoI performance of a multi-queue remote monitoring system where several devices observe the same physical process and transmit status updates to the monitor.
Two types of devices are considered.
The type-1 devices take the ZW policy for sampling and follow exponentially distributed service times.
The type-2 devices transmit Poisson arrival updates instantaneously.
According to whether there are type-2 devices in the system, we consider two multi-queue systems, namely the \emph{Basic System} where only type-1 devices are equipped and the \emph{Hybrid System} where both type-1 devices and type-2 devices are equipped.
By leveraging the SHS method, we derive the closed-form expressions of the average AoI for the two systems.
Numerical results validate our analysis.
It is shown that the logarithm of average AoI decreases linearly with the logarithm of the number of the type-1 devices in \emph{Basic System} and the total arrival rate of the type-2 devices in \emph{Hybrid System} when the service rate of all type-1 devices are set to 0.
It is also shown that the benefit brought by increasing unit device is lower for type-1 device compared to type-2 device.
In addition, for practical deployment, it is found that there is a threshold interval for the total cost of deployment units to determine the optimal deployment scheme.
The analysis of our model further expands the timeliness analysis theory of the AoI based network queuing modeling system.
The future work is to consider cascading two types of devices, or further considering multiple devices with general arrival rates and general service rates.
\appendix
\subsection{Proof of Theorem 1}\label{Theorem_1_proof}
To compute the average AoI of the considered system, let us specify all the SHS elements, including the parameters of all transitions such as transition rate, reset map, and differential equations describing the evolution of $\boldsymbol{x}$.
\begin{table*}[t]
\centering
\caption{SHS Elements Corresponding to A Typical  Transition $l\in\mathcal{L}$.}
\label{Tab_l_nstream}
\begin{tabular}{ccccc}
\toprule
Case & Refresh & $\lambda^{(l)}$ & $\boldsymbol{x}'=\boldsymbol{x}\boldsymbol{A}_{l}$ & $\boldsymbol{v}'_{\boldsymbol{q}_{l}}=\boldsymbol{v}_{\boldsymbol{q}_{l}}\boldsymbol{A}_{l}$\\
\midrule
\specialrule{0em}{1pt}{1pt}

\begin{IEEEeqnarraybox}{s.t}
$q'_{l}(1)=$
$q_{l}(p(\boldsymbol{q}'_{l})-1)$
\end{IEEEeqnarraybox}
&
YES
&
$\mu_{q'_{l}(1)}$
&
\multicolumn{1}{l}{
$x'_{i}=
\begin{cases}
x_{q'_{l}(1)},&\;\;\! i=0,\\
0,&\;\;\! i=q'_{l}(1),\\
x_{i},&\;\;\!\text{otherwise}.
\end{cases}
$}
&
$v'_{\boldsymbol{q}_{l}i}=
\begin{cases}
v_{\boldsymbol{q}_{l}q'_{l}(1)},&i=0,\\
0,&i=q'_{l}(1),\\
v_{\boldsymbol{q}_{l}i},&\text{otherwise}.
\end{cases}
$
\\
\specialrule{0em}{1pt}{1pt}
\multicolumn{1}{l}{
\begin{IEEEeqnarraybox}{s}
$q'_{l}(1)\neq$
$q_{l}(p(\boldsymbol{q}'_{l})-1)$
\end{IEEEeqnarraybox}
}
&
NO
&
$\mu_{q'_{l}(1)}$
&
\multicolumn{1}{l}{
$x'_{i}=
\begin{cases}
0,    &  \quad \;\,\: i=q'_{l}(1),\\
x_{i},&  \quad \;\,\: \text{otherwise}.
\end{cases}
$}
&
\multicolumn{1}{l}{
$ v'_{\boldsymbol{q}_{l}i}=
\begin{cases}
0,& \quad\;\:i=q'_{l}(1),\\
v_{\boldsymbol{q}_{l}i},& \quad\;\: \text{otherwise}.
\end{cases}
$}
\\
\bottomrule
\end{tabular}
\vspace{-1em}
\end{table*}

We start by clarifying the transition rate of transition $l~(l\in\mathcal{L})$.
By definition, after a transition $l$: $\boldsymbol{q}_{l}\to \boldsymbol{q}'_{l}$, the latest $(N-1)$ updates before the transition would become those from the second latest to the $N$-th latest updates, i.e.,
\begin{equation}\label{ql_to_ql'}
q'_{l}(i)=q_{l}(i-1), \quad i\in \{2,~3,~\cdots,~N\}.
\end{equation}
Accordingly, the transition rate of transition $l$, which is denoted by $\lambda^{(l)}$, is $\mu_{q'_{l}(1)}$.
In general, for any $\boldsymbol{q}\in \mathcal{Q}$,
\begin{equation}\label{lambda_l}
    \lambda^{(l)}=\mu_{q(1)},~\forall l\in \mathcal{L}'_{\boldsymbol{q}}.
\end{equation}

Then, to get the reset map of transition $l$, it is equivalent to analyzing the change of $\boldsymbol{x}$ when transition $l$ takes place. Since the latest update comes from type-1 device $q'_{l}(1)$, we assert that the AoI of update in type-1 device $q'_{l}(1)$ would be reset to $0$ while the AoI of update in other type-1 devices remains the same.
For the AoI of the monitor, there are two likely outcomes.
If $x_{q'_{l}(1)}<x_{0}$, then it would be refreshed to $x_{q'_{l}(1)}$, i.e., $x'_{0}=x_{q'_{l}(1)}$, otherwise $x'_{0}=x_{0}$.
With the help of the function $p(\boldsymbol{q})$, let us explain how to decide whether the AoI of the monitor would be refreshed or not based on two cases in Table \ref{Tab_l_nstream} in detail.

\begin{figure}[t]
\centering
\subfigure[Case $1$: $q'_{l}(1)=q_{l}(r)$, where $r=p(\boldsymbol{q}'_{l})-1$.]
{\label{condition1}\includegraphics[width=225pt]{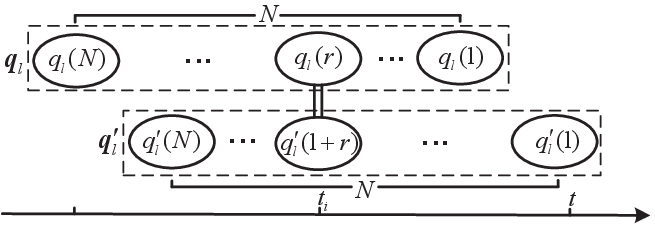}} 
\subfigure[Case $2$: $q'_{l}(1)\neq q_{l}(r)$, where $r=p(\boldsymbol{q}'_{l})-1$.]  
{\label{condition2}\includegraphics[width=225pt]{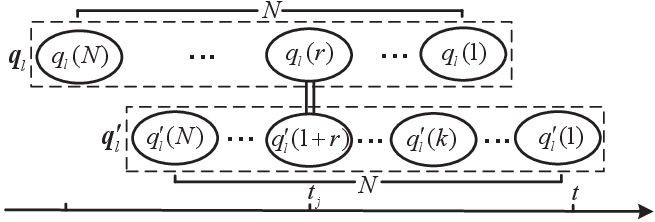}}
\caption{The two cases of $l$ in Tab. \ref{Tab_l_nstream}.} 
\label{condition}  
\vspace{-1em}
\end{figure}
\begin{itemize}
  \item Case $1$: $q'_{l}(1)=q_{l}(p(\boldsymbol{q}'_{l})-1)$. This case is illustrated in Fig.~\ref{condition1} where $r=p(\boldsymbol{q}'_{l})-1$ and $t_{i}$ is the arrival time of the $(1+r)$-th latest update which is assumed to come from type-1 device $i$ $(i\in \mathcal{N})$. The latest arrived update arrives at time $t$ and comes from type-1 device $i$ since $q'_{l}(1)=q_{l}(r)=q'_{l}(1+r)$.
      Thus, the generation time of the latest arrived update is $t_{i}$.
  Then, according to the definition of $p(\boldsymbol{q})$, the source of all updates arrived at the monitor between $t_{i}$ and $t$ are different.
  Thus we assert that the generation time of the latest updates from all other type-1 devices is earlier than $t_{i}$, i.e., the AoI of the latest arrived update is smaller than that of all the latest updates coming from other type-1 devices.
  Thus we assert that $x_{i}(t)<x_{0}(t)$ which implies that the AoI of the monitor would be refreshed.

  \item  Case $2$: $q'_{l}(1)\neq q_{l}(p(\boldsymbol{q}'_{l})-1)$. This case is illustrated in Fig.~\ref{condition2} where $r=p(\boldsymbol{q}'_{l})-1$ and $t_{j}$ is the arrival time of the $(1+r)$-th latest update which is assumed to come from type-1 device $j$ ($j\in \mathcal{N}$), i.e., $q'_{l}(r+1)=q_{l}(r)=j$.
      The latest arrived update arrives at time $t$ and comes from type-1 device $q'_{l}(1)$. Without loss of generality, let us assume that $q'_{l}(1)=i$ $(i\in\mathcal{N},i\neq j)$. Since $q'_{l}(1)\neq q_{l}(r)=q'_{l}(1+r)$, there must exist a unique $k$ $(k\in\{2,3,\cdots,r\})$ such that $q'_{l}(k)=q'_{l}(1+r)$, i.e., the $k$-th latest update comes from the same type-1 device as the $(1+r)$-th latest update.
      Thus, one can find that the generation time of the latest arrived update is earlier than $t_{j}$, i.e., the generation time of $k$-th latest update, which implies that the AoI of the latest arrived update is greater than that of the $k$-th latest update. Thus one can deduce that $x_{0}(t)<x_{i}(t)$ and the AoI of the monitor would not be refreshed.
\end{itemize}

With the knowledge of the change from $\boldsymbol{x}$ to $\boldsymbol{x}'$, one can express $\boldsymbol{A}_{l}$ explicitly and further deduce the change from $\boldsymbol{v}_{\boldsymbol{q}_{l}}$ to $\boldsymbol{v}'_{\boldsymbol{q}_{l}}=[v'_{\boldsymbol{q}_{l}0}\quad  \cdots \quad v'_{\boldsymbol{q}_{l}N}]$ as shown in Table \ref{Tab_l_nstream}.

Next, we further analyze the differential equations used in the evolution of $\boldsymbol{x}$.
Since the \emph{Basic System} utilizes the ZW policy, $x_{i}$ $(i\in \{0\} \cup \mathcal{N})$ always increases at a unit rate, i.e.,
\begin{equation}\label{nbq}
\boldsymbol{b}_{\boldsymbol{q}}=[\underbrace{1~\cdots~ 1}_{N+1}],\quad \boldsymbol{q}\in \mathcal{Q}.
\end{equation}

With the analytic expression of the SHS elements, now one can derive the average AoI by finding $\boldsymbol{v}_{\boldsymbol{q}}$ through \eqref{correlation vector}.
Before that, it requires to find the steady state probability vector $\overline{\pi}_{\boldsymbol{q}}$ $(\boldsymbol{q}\in \mathcal{Q})$ according to \eqref{steady state probability1} and \eqref{steady state probability2}.
Since for any $\boldsymbol{q}\in \mathcal{Q}$, there are $N$ outgoing transitions whose transition rates are $\mu_{1}$ to
$\mu_{N}$ respectively, it has
\begin{equation}\label{sum_lambda_l}
 \sum_{l\in\mathcal{L}_{\boldsymbol{q}}}\lambda^{(l)}=\sum_{i=1}^{N}\mu_{i}=\mu.
\end{equation}
Combining \eqref{lambda_l} and \eqref{sum_lambda_l}, \eqref{steady state probability1} can be simplified as
\begin{equation}\label{steady state probability1_sim}
  \overline{\pi}_{\boldsymbol{q}}=\frac{\mu_{q(1)}}{\mu}\sum_{l\in\mathcal{L}_{\boldsymbol{q}}}\overline{\pi}_{\boldsymbol{q}_{l}}.
\end{equation}
Define $\mathcal{Q}_{\boldsymbol{q}}^{(i)}=\{\boldsymbol{q}'\in\mathcal{Q}:q'(j)=q(j+i),\forall j\in\{1,\cdots,N-i\}\}$ as the set of possible states which can transit to the current state through $i$ times of transitions.
For instance, one can see that $\mathcal{Q}_{\boldsymbol{q}}^{(1)}=\{\boldsymbol{q}_{l}\in \mathcal{Q}:l \in \mathcal{L}'_{\boldsymbol{q}}\}$ and $\mathcal{Q}_{\boldsymbol{q}}^{(N)}=\mathcal{Q}$.
Then, $\overline{\pi}_{\boldsymbol{q}}$ can be found as
\begin{align}\label{nstream_steady state probability}
\overline{\pi}_{\boldsymbol{q}}&
             =\frac{\mu_{q(1)}}{\mu}\sum_{\boldsymbol{q}'\in \mathcal{Q}_{\boldsymbol{q}}^{(1)}}\overline{\pi}_{\boldsymbol{q}'}
             =\frac{\mu_{q(1)}\mu_{q(2)}}{\mu^2}\sum_{\boldsymbol{q}'\in \mathcal{Q}_{\boldsymbol{q}}^{(2)}}\overline{\pi}_{\boldsymbol{q}'}
             \notag\\
             &
             \stackrel{\text{(a)}}{=}\frac{M(\boldsymbol{q})}{\mu^{N}}\sum_{\boldsymbol{q}'\in\mathcal{Q}_{\boldsymbol{q}}^{(N)}}\overline{\pi}_{\boldsymbol{q}'}
             =\frac{M(\boldsymbol{q})}{\mu^{N}}\sum_{\boldsymbol{q}'\in\mathcal{Q}}\overline{\pi}_{\boldsymbol{q}'}
             \stackrel{\text{(b)}}{=}\frac{M(\boldsymbol{q})}{\mu^{N}},
\end{align}
where ${\text{(a)}}$ follows from \eqref{steady state probability1_sim}, and ${\text{(b)}}$ follows from \eqref{steady state probability2}.

So far, it remains to get $\overline{\boldsymbol{v}}_{\boldsymbol{q}}$ by \eqref{correlation vector}. In fact, it is equivalent to solving $\overline{v}_{\boldsymbol{q}i}$ through \eqref{stable vqi_sim} which is obtained from \eqref{correlation vector} using \eqref{lambda_l}, \eqref{nbq}, \eqref{sum_lambda_l}, and \eqref{nstream_steady state probability}.
\begin{equation}\label{stable vqi_sim}
  \overline{v}_{\boldsymbol{q}i} = \frac{M(\boldsymbol{q})}{\mu^{N+1}}+\frac{\mu_{q(1)}}{\mu}\sum_{l\in \mathcal{L}'_{\boldsymbol{q}}}\overline{v}'_{\boldsymbol{q}_{l}i}, \boldsymbol{q}\in\mathcal{Q},i\in \{0\}\cup\mathcal{N}.
\end{equation}

Note that what we indeed concern about is $\overline{v}_{\boldsymbol{q}0}$.
From \eqref{stable vqi_sim}, we can find that $\overline{v}_{\boldsymbol{q}0}$ is related to $\overline{v}'_{\boldsymbol{q}_{l}0}$ $(l\in\mathcal{L}'_{\boldsymbol{q}})$ which probably takes the value of $\overline{v}_{\boldsymbol{q}_{l}q(1)}$ or $\overline{v}_{\boldsymbol{q}_{l}0}$ based on the two specific cases considered in Tab. \ref{Tab_l_nstream}.
In particular, when $\overline{v}'_{\boldsymbol{q}_{l}0}=\overline{v}_{\boldsymbol{q}_{l}q(1)}$, it must have $q(1)=q_{l}(p(\boldsymbol{q})-1)$.
It means that if $\overline{v}_{\boldsymbol{q}_{l}i}$ is the term that a $\overline{v}_{\boldsymbol{q}0}$ might be related to, then it must satisfy the condition $i\in \mathcal{N}_{\boldsymbol{q}_{l}}$ where $\mathcal{N}_{\boldsymbol{q}}=\{i\in\mathcal{N}:\exists j\in\mathcal{N},q(j)=i\}$.
Thus for $\boldsymbol{q}\in \mathcal{Q}$, let us first solve $\overline{v}_{\boldsymbol{q}i}$ ($i\in\mathcal{N}_{\boldsymbol{q}}$), and then derive $\overline{v}_{\boldsymbol{q}0}$ based on $\overline{v}_{\boldsymbol{q}i}$.

To derive $\overline{v}_{\boldsymbol{q}i}$ ($i\in\mathcal{N}_{\boldsymbol{q}}$), let us define a function $p_{i}(\boldsymbol{q})=\min\limits_{r\in\{\overline{r}\in\mathcal{N}:q(\overline{r})=i\}}r$ for a state $\boldsymbol{q}$ which indicates the index of the latest update that comes from type-1 device $i$.
Note that $p_{i}(\boldsymbol{q})\in\mathcal{N}$. Based on the value of $p_{i}(\boldsymbol{q})$, we can get $\overline{v}_{\boldsymbol{q}i}$ as follows.
\begin{itemize}
\item $p_{i}(\boldsymbol{q})=1$: In this case, one can find that $\overline{v}'_{\boldsymbol{q}_{l}i}=0$ for all $l\in\mathcal{L}'_{\boldsymbol{q}}$ (cf. Tab. \ref{Tab_l_nstream}) since $q'_{l}(1)=q(1)=i$, then one can get $\overline{v}_{\boldsymbol{q}i}$ from \eqref{stable vqi_sim}
    \begin{equation}\label{vqi_1}
      \overline{v}_{\boldsymbol{q}i}=\frac{M(\boldsymbol{q})}{\mu^{N+1}}.
    \end{equation}
\item  $2\le p_{i}(\boldsymbol{q})\le N$: In this case, one can find that $\overline{v}'_{\boldsymbol{q}_{l}i}=\overline{v}_{\boldsymbol{q}_{l}i}$ for all $l\in\mathcal{L}'_{\boldsymbol{q}}$ (cf. Tab. \ref{Tab_l_nstream}) since $ q'_{l}(1)\neq i$ and $i\neq 0$. Then according to \eqref{stable vqi_sim}, one has that
    \begin{equation}\label{vqi2}
    \overline{v}_{\boldsymbol{q}i}=\frac{M(\boldsymbol{q})}{\mu^{N+1}}+\frac{\mu_{q(1)}}{\mu}\sum_{l\in \mathcal{L}'_{\boldsymbol{q}}}\overline{v}_{\boldsymbol{q}_{l}i}.
    \end{equation}
It can be found that $p_{i}(\boldsymbol{q}_{l})=p_{i}(\boldsymbol{q})-1$ $(l\in \mathcal{L}'_{\boldsymbol{q}})$ from \eqref{ql_to_ql'}. Then one can get $\overline{v}_{\boldsymbol{q}i}$ by recursion as follows
\begin{align}\label{vqi_n}
\overline{v}_{\boldsymbol{q}i}={}&\frac{M(\boldsymbol{q})}{\mu^{N+1}}+\frac{\mu_{q(1)}}{\mu}\sum_ {\tilde{\boldsymbol{q}}\in \mathcal{Q}_{\boldsymbol{q}}^{(1)}}\Bigl(\frac{M(\tilde{\boldsymbol{q}})} {\mu^{N+1}}
                              +\frac{\mu_{\tilde{q}(1)}}{\mu}\sum_{l\in\mathcal{L}'_{\tilde{\boldsymbol{q}}}}\overline{v}_{\tilde{\boldsymbol{q}}_{l}i}\Bigl)
                              \notag\\
                              ={}&
                              \frac{2M(\boldsymbol{q})}{\mu^{N+1}}+\frac{\mu_{q(1)}\mu_{q(2)}}{\mu^{2}}\sum_ {\tilde{\boldsymbol{q}}\in \mathcal{Q}_{\boldsymbol{q}}^{(2)}}\overline{v}_{\tilde{\boldsymbol{q}}i}\notag\\
                              \stackrel{\text{(c)}}{=}{}&\frac{(p_{i}(\boldsymbol{q})-1)M(\boldsymbol{q})}{\mu^{N+1}}
                              +\frac{\prod_{j=1}^{p_{i}(\boldsymbol{q})-1}\mu_{q(j)}}{\mu^{(p_{i}(\boldsymbol{q})-1)}}\!\!\sum_ {\tilde{\boldsymbol{q}}\in \mathcal{Q}_{\boldsymbol{q}}^{(p_{i}(\boldsymbol{q})-1)}}\overline{v}_{\tilde{\boldsymbol{q}}i}\notag\\
                             \stackrel{\text{(d)}}{=}{}&\frac{(p_{i}(\boldsymbol{q})-1)M(\boldsymbol{q})}{\mu^{N+1}}
                             \notag\\
                             &
                             +\frac{\prod_{j=1}^{p_{i}(\boldsymbol{q})-1}\mu_{q(j)}}{\mu^{(p_{i}(\boldsymbol{q})-1)}}\cdot \frac{\mu^{p_{i}(\boldsymbol{q})-1}M(\boldsymbol{q})}{\prod_{j=1}^{p_{i}(\boldsymbol{q})-1}\mu_{q(j)}\mu^{N+1}}\notag\\
                             ={}&\frac{p_{i}(\boldsymbol{q})M(\boldsymbol{q})}{\mu^{N+1}},
\end{align}
where ${\text{(c)}}$ follows from \eqref{vqi2}, and $\text{(d)}$ follows from \eqref{vqi_1}.
\end{itemize}

Combining \eqref{vqi_1} and \eqref{vqi_n}, for any $\boldsymbol{q}\in \mathcal{Q},i\in \mathcal{N}_{\boldsymbol{q}}$, it has
\begin{equation}\label{vqn}
\overline{v}_{\boldsymbol{q}i}=\frac{p_{i}(\boldsymbol{q})M(\boldsymbol{q})}{\mu^{N+1}}.
\end{equation}

Finally, according to the value of $p(\boldsymbol{q})$ and whether type-1 device $q(1)$ is the the source of the update repeats in the previous updates for the first time, one can deduce $\overline{v}_{\boldsymbol{q}0}$ as follows.

\begin{table*}[t]
\centering
\caption{Parameters of A Typical Transition $l$}
\label{Tab_l_n_1stream}
\begin{tabular}{ccccc}
\toprule
Case & Refresh & $\lambda^{(l)}$ & $\boldsymbol{x}'=\boldsymbol{x}\boldsymbol{A}_{l}$ & $\boldsymbol{v}'_{\boldsymbol{q}_{l}}=\boldsymbol{v}_{\boldsymbol{q}_{l}}\boldsymbol{A}_{l}$\\
\midrule
\specialrule{0em}{1pt}{1pt}
$\tilde{p}(\boldsymbol{q}'_{l})=1$
&
YES
&
$\lambda$
&
$x'_{i}=
\begin{cases}
0,    &\;\;\, i=0,\\
x_{i},&\;\;\, \text{otherwise}.
\end{cases}
$
&
$ v'_{\boldsymbol{q}_{l}i}=
\begin{cases}
0,                      & \quad i=0,\\
v_{\boldsymbol{q}_{l}i},& \quad \text{otherwise}.
\end{cases}
$
\\
\specialrule{0em}{1pt}{1pt}
$
1<\tilde{p}(\boldsymbol{q}'_{l})<p(\boldsymbol{q}'_{l})
$
&
NO
&
$u_{q'_{l}(1)}$
&
\multicolumn{1}{l}{
$x'_{i}=
\begin{cases}
0,    &\;\;\,  i=q'_{l}(1),\\
x_{i},&\;\;\,  \text{otherwise}.
\end{cases}
$}
&
\multicolumn{1}{l}{
$ v'_{\boldsymbol{q}_{l}i}=
\begin{cases}
0,                      & \quad i=q'_{l}(1),\\
v_{\boldsymbol{q}_{l}i},& \quad \text{otherwise}.
\end{cases}
$}
\\
\specialrule{0em}{1pt}{1pt}

$
\begin{cases}
\tilde{p}(\boldsymbol{q}'_{l})\geq p(\boldsymbol{q}'_{l})\\
q'_{l}(1)=
q_{l}(p(\boldsymbol{q}'_{l})-1)
\end{cases}
$

&
YES
&
$u_{q'_{l}(1)}$
&
\multicolumn{1}{l}{
$x'_{i}=
\begin{cases}
x_{q'_{l}(1)},& \!\!\!\!\!\: i=0,\\
0,&             \!\!\!\!\!\: i=q'_{l}(1),\\
x_{i},&         \!\!\!\!\!\: \text{otherwise}.
\end{cases}
$}
&
$v'_{\boldsymbol{q}_{l}i}=
\begin{cases}
v_{\boldsymbol{q}_{l}q'_{l}(1)},& \!\!\! i=0,\\
0,                              & \!\!\! i=q'_{l}(1),\\
v_{\boldsymbol{q}_{l}i},        & \!\!\! \text{otherwise}.
\end{cases}
$
\\
\specialrule{0em}{1pt}{1pt}
$
\begin{cases}
\tilde{p}(\boldsymbol{q}'_{l})\geq p(\boldsymbol{q}'_{l})\\
q'_{l}(1)\neq
q_{l}(p(\boldsymbol{q}'_{l})-1)
\end{cases}
$
&
NO
&
$u_{q'_{l}(1)}$
&
\multicolumn{1}{l}{
$x'_{i}=
\begin{cases}
0,    & \;\;\, i=q'_{l}(1),\\
x_{i},& \;\;\, \text{otherwise}.
\end{cases}
$}
&
\multicolumn{1}{l}{
$ v'_{\boldsymbol{q}_{l}i}=
\begin{cases}
0,&                       \quad i=q'_{l}(1),\\
v_{\boldsymbol{q}_{l}i},& \quad \text{otherwise}.
\end{cases}
$}
\\
\bottomrule
\end{tabular}
\vspace{-1em}
\end{table*}

\begin{itemize}
  \item $2\le p(\boldsymbol{q})\le N$ and $q(1)=q(p(\boldsymbol{q}))$: In this case, one can find that $\overline{v}'_{\boldsymbol{q}_{l}0}=\overline{v}_{\boldsymbol{q}_{l}q(1)}$ for all $l\in\mathcal{L}'_{\boldsymbol{q}}$ (cf. Tab. \ref{Tab_l_nstream}) since $q'_{l}(1)=q'_{l}(p(\boldsymbol{q}'_{l}))=q_{l}(p(\boldsymbol{q}'_{l})-1)$. One can also find $p_{q(1)}(\boldsymbol{q}_{l})=p(\boldsymbol{q})-1$ $(l\in\mathcal{L}'_{\boldsymbol{q}})$ from \eqref{ql_to_ql'}. Then, we can get $\overline{v}_{\boldsymbol{q}0}$ from \eqref{stable vqi_sim} and \eqref{vqn} as
      \begin{align}\label{vq0_1}
        \overline{v}_{\boldsymbol{q}0} ={} &  \frac{M(\boldsymbol{q})}{\mu^{(N+1)}}+\frac{\mu_{q(1)}}{\mu}\sum_{l\in \mathcal{L}'_{\boldsymbol{q}}}\overline{v}_{\boldsymbol{q}_{l}q(1)}
        \notag\\
                                       \stackrel{\text{(e)}}{=}{} & \frac{M(\boldsymbol{q})}{\mu^{N+1}}
                                       +\frac{\mu_{q(1)}}{\mu}\frac{(p(\boldsymbol{q})-1)M(\boldsymbol{q})/
                                       \mu_{q(1)}}{\mu^{N+1}}\mu
                                       \notag\\
                                       ={}&
                                       \frac{p(\boldsymbol{q})M(\boldsymbol{q})}{\mu^{N+1}},
      \end{align}
      where $\text{(e)}$ follows from \eqref{vqn}.
  \item $2\le p(\boldsymbol{q})\le N$ and $q(1) \neq q(p(\boldsymbol{q}))$: In this case, one can find that $\overline{v}'_{\boldsymbol{q}_{l}0}=\overline{v}_{\boldsymbol{q}_{l}0}$ (cf. Tab. \ref{Tab_l_nstream}) for all $l\in\mathcal{L}'_{\boldsymbol{q}}$ since $q'_{l}(1)\neq q_{l}(p(\boldsymbol{q}'_{l})-1)$. Then based on \eqref{stable vqi_sim}, it has
      \begin{align}\label{vq02}
        \overline{v}_{\boldsymbol{q}0}  &= \frac{M(\boldsymbol{q})}{\mu^{(N+1)}}+\frac{\mu_{q(1)}}{\mu}\sum_{l\in \mathcal{L}'_{\boldsymbol{q}}}\overline{v}_{\boldsymbol{q}_{l}0}.
      \end{align}
      It can also be found that $p(\boldsymbol{q}_{l})=p(\boldsymbol{q})-1$ $(l\in\mathcal{L}'_{\boldsymbol{q}})$ from \eqref{ql_to_ql'}.
      Since $q(1) \neq q(p(\boldsymbol{q}))$, there must exist a unique $k$ ($k\in \{2,\cdots,p(\boldsymbol{q})-1\}$) such that $q(k)=q(p(\boldsymbol{q}))$.
      Then, one can deduce $\overline{v}_{\boldsymbol{q}0}$ using a recursive method similar to that used in \eqref{vqi_n}, i.e.,
        \begin{align}\label{vq0_2}
         \overline{v}_{\boldsymbol{q}0}{} 
                                        \stackrel{\text{(f)}}{=}{}&\frac{(k-1)M(\boldsymbol{q})}{\mu^{N+1}}
                                        +\frac{\prod_{j=1}^{k-1}\mu_{q(j)}}{\mu^{k-1}}\sum_ {\tilde{\boldsymbol{q}}\in \mathcal{Q}_{\boldsymbol{q}}^{(k-1)}}\overline{v}_{\tilde{\boldsymbol{q}}0}\notag\\
                                        \stackrel{\text{(g)}}{=}{}&\frac{(k-1)M(\boldsymbol{q})}{\mu^{N+1}}
                                        \notag\\
                                        &
                                        +\frac{\prod_{j=1}^{k-1}\mu_{q(j)}}{\mu^{k-1}}
                                        \frac{(p(\boldsymbol{q})-k+1)M(\boldsymbol{q})}{\prod_{j=1}^{k-1}\mu_{q(j)}\mu^{N+1}}\mu^{k-1}\notag\\
                                        ={}&\frac{p(\boldsymbol{q})M(\boldsymbol{q})}{\mu^{N+1}},
      \end{align}
  where $\text{(f)}$ follows from by \eqref{vq02}, and $\text{(g)}$ follows from \eqref{vq0_1}.
  \item $p(\boldsymbol{q})=N+1$: In this case, one can find that for $l\in\mathcal{L}'_{\boldsymbol{q}}$, $\overline{v}'_{\boldsymbol{q}_{l}0}=\overline{v}_{\boldsymbol{q}_{l}q(1)}$ if $q_{l}(N)=q(1)$, otherwise $\overline{v}'_{\boldsymbol{q}_{l}0}= \overline{v}_{\boldsymbol{q}_{l}0}$ (cf. Tab. \ref{Tab_l_nstream}).
   In fact, there is only one transition $l$ such that $q_{l}(N)=q(1)$. Denote this transition by $l_{0}$. Then, $q_{l_{0}}(N)=q(1)$. Thus, one can find that $p_{q(1)}(\boldsymbol{q}_{l_{0}})=N$ and $p(\boldsymbol{q}_{l})=N$ $(l\in \mathcal{L}'_{\boldsymbol{q}},l \neq l_{0})$.
   Then one can get 
      \begin{align}\label{vq0_4}
        \overline{v}_{\boldsymbol{q}0}={} & \frac{M(\boldsymbol{q})}{\mu^{N+1}}+\frac{\mu_{q(1)}}{\mu}\sum_{\substack{l\in \mathcal{L}'_{\boldsymbol{q}},l\ne l_{0}}}\overline{v}_{\boldsymbol{q}_{l}0}+
                                            \frac{\mu_{q(1)}}{\mu}\overline{v}_{\boldsymbol{q}_{l_{0}}q(1)}  \notag\\
                                      \stackrel{\text{(h)}}{=}{} & \frac{M(\boldsymbol{q})}{\mu^{N+1}}+\frac{\mu_{q(1)}}{\mu}\frac{NM(\boldsymbol{q})/\mu_{q(1)}}
                                            {\mu^{N+1}}\mu
                                            \notag\\
                                          &
                                          -\frac{\mu_{q(1)}}{\mu}\frac{NM(\boldsymbol{q})/\mu_{q(1)}}
                                            {\mu^{N+1}}\mu_{q_{l_{0}}(N)}
                                           +\frac{\mu_{q(1)}}{\mu}\frac{NM(\boldsymbol{q})}{\mu^{N+1}} \notag\\
                                      ={} & \frac{(N+1)M(\boldsymbol{q})}{\mu^{N+1}},
      \end{align}
      where $\text{(h)}$ follows from \eqref{vq0_1}, \eqref{vq0_2}, and \eqref{vqn} .
\end{itemize}

Combining \eqref{vq0_1}, \eqref{vq0_2} and \eqref{vq0_4}, it has that for $\boldsymbol{q}\in \mathcal{Q}$,
\begin{equation}\label{vq0}
  \overline{v}_{\boldsymbol{q}0}=\frac{p(\boldsymbol{q})M(\boldsymbol{q})}{\mu^{N+1}}.
\end{equation}
Further using \eqref{vq0} and \eqref{average AoI}, one can get ${\Delta}_{N}^{\text{Basic}}$ as \eqref{nstreamaverageAoI}.

\subsection{Proof of Corollary 1}\label{Corollary_1_proof}
Recall that $2\leq p(\boldsymbol{q})\leq N+1$. By substituting $\mu_{i}={\mu}/{N}$ into \eqref{nstreamaverageAoI}, it has
\begin{equation}\label{nstreamaverageAoI_s1}
  \Delta _{N}^{\text{Bas-Homo}}=\sum_{\boldsymbol{q}\in \mathcal{Q}}\frac{p(\boldsymbol{q})(\mu/N)^{N}}{\mu^{N+1}}=\sum_{r=1}^{N}\frac{C(r)(1+r)}{N^{N}\mu},
\end{equation}
where $C(r)$ represents the number of $\boldsymbol{q}$ satisfying $p(\boldsymbol{q})=1+r$.
According to the value of $r$ ($1\leq r \leq N$), one can derive $C(r)$ as follows.
\begin{itemize}
  \item $r=1$: In this case, the second latest to the $N$-th latest update can come from any type-1 device and the latest update comes from the same device as the second latest update, thus $C(1)=N^{N-1}$.
  \item $1<r<N$: In this case, the ($N-r$) updates from $(1+r)$-th latest to the $N$-th latest can come from any type-1 device, while the latest $r$ updates come from different type-1 devices and one of the latest $r$ updates comes from the same sensor as the $(1+r)$-th latest update.
      Accordingly, it has
      \begin{equation}\label{C(r)}
        C(r)=N^{(N-r)}\binom{N-1}{r-1}r!.
      \end{equation}
  \item $r=N$: In this case, the latest $N$ updates come from different type-1 devices, thus $C(N)=N!$.
\end{itemize}
It can be found that $C(1)$ and $C(N)$ also satisfy \eqref{C(r)}.
Plugging \eqref{C(r)} into \eqref{nstreamaverageAoI_s1}, one can get $\Delta _{N}^{\text{Bas-Homo}}$ as \eqref{corollary_1}.
\vspace{-0.5em}

\subsection{Proof of Theorem 2}\label{Theorem_2_proof}
To compute the average AoI, we still start with specifying all the SHS elements.
First, let us clarify the parameters of all transitions, including transition rate and reset map.
For a transition $l\in\mathcal{L}$, we also have \eqref{ql_to_ql'}. Accordingly, the transition rate of transition $l$ is $\lambda^{(l)}=u_{q'_{l}(1)}$.
In general, for any $\boldsymbol{q}\in \mathcal{Q}$,
\begin{equation}\label{lambda_l2}
    \lambda^{(l)}=u_{q(1)},~\forall l\in \mathcal{L}'_{\boldsymbol{q}}.
\end{equation}
Then, to get its reset map, it needs to know the change of $\boldsymbol{x}$ when transition $l$ takes place.
Since the latest update would only reset the age of its source device to $0$ without affecting the age of other devices, we just need to focus on the change of the age of the monitor.
With the help of functions $p(\boldsymbol{q})$ and $\tilde{p}(\boldsymbol{q})$, we give the variation of $\boldsymbol{x}$ under the four cases for transition $l$ as shown in Table \ref{Tab_l_n_1stream}.
In the following, let us explain how to decide whether the AoI of the monitor would be refreshed or not based on the four specific cases.

\begin{figure}[t]
\centering
\subfigure[Case $2$: $1 < d < r$, where $d = \tilde{p}(\boldsymbol{q}'_{l})$, $r = p(\boldsymbol{q}'_{l})$.] 
{\label{case2}\includegraphics[width=220pt]{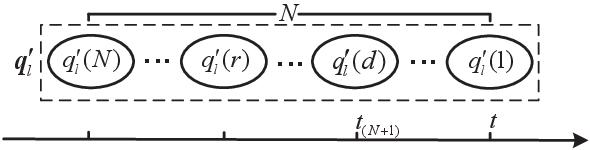}} 
\subfigure[Case $3$: $d \geq 1+r$ and $q'_{l}(1)=q_{l}(r)$, where $d = \tilde{p}(\boldsymbol{q}'_{l})$, $r = p(\boldsymbol{q}'_{l})-1$.]
{\label{case3}\includegraphics[width=220pt]{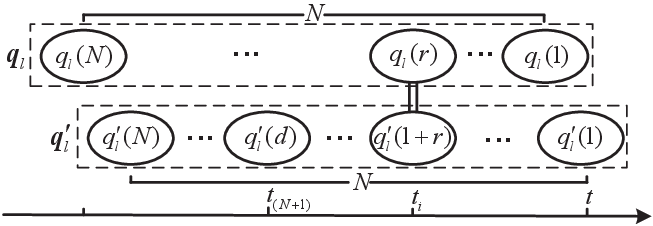}}

\subfigure[Case $4$: $d \geq 1+r$ and $q'_{l}(1) \neq q_{l}(r)$, where $d = \tilde{p}(\boldsymbol{q}'_{l})$, $r = p(\boldsymbol{q}'_{l})-1$.]
{\label{case4}\includegraphics[width=220pt]{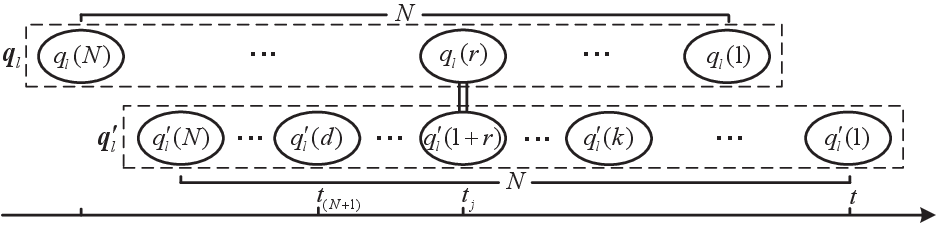}}
\caption{The second to fourth cases of $l$ in Tab. \ref{Tab_l_n_1stream}.} 
\label{Sg_Case}  
\vspace{-1.5em}
\end{figure}
\begin{itemize}
    \item Case $1$: $\tilde{p}(\boldsymbol{q}'_{l})=1$. In this case, the age of the monitor will obviously be refreshed to $0$, i.e., $x'_{0}=0$, since the latest arrived update comes from device $(N+1)$.
    \item Case $2$: $1<\tilde{p}(\boldsymbol{q}'_{l})<p(\boldsymbol{q}'_{l})$. This is described in Fig.~\ref{case2} where $r=p(\boldsymbol{q}'_{l}),~d=\tilde{p}(\boldsymbol{q}'_{l})$.
        The latest arrived update that arrives at time $t$ is assumed to come from device $i~(i\in \mathcal{N})$.
        The $d$-th latest update is the latest update that comes from device $(N+1)$.
        Both its generation time and arrival time are $t_{(N+1)}$.
        By definition, the sources of updates that arrive after the $r$-th latest update are different.
        Furthermore, according to the condition, one can know that all updates that arrive after $t_{(N+1)}$ come from different devices.
        This means that the generation time of the latest arrived update is earlier than $t_{(N+1)}$, i.e., the AoI of the latest arrived update is greater than the AoI of the $r$-th latest update. Thus the AoI of the monitor would remain the same.
    \item Case $3$: $\tilde{p}(\boldsymbol{q}'_{l})\geq p(\boldsymbol{q}'_{l})$ and $q'_{l}(1)=q_{l}(p(\boldsymbol{q}'_{l})-1)$. This is  described in Fig.~\ref{case3} where $d = \tilde{p}(\boldsymbol{q}'_{l})$ and $r=p(\boldsymbol{q}'_{l})-1$.
        The latest arrived update comes from the same device as the $(1+r)$-th latest update and the arrival time of the latest update from device $(N+1)$ $t_{N+1}$ is earlier than that of the $(1+r)$-th latest update $t_{i}$.
        Then the situation is similar to Case $1$ in the \emph{Basic System}.
        Since all updates arrived between $t_{i}$ and $t$ are not from device $(N+1)$ and they come from different devices, we assert that the AoI of the latest arrived update is smaller than that of all the latest updates from other devices. Accordingly, the AoI of the monitor would be refreshed.
    \item Case $4$: $\tilde{p}(\boldsymbol{q}'_{l})\geq p(\boldsymbol{q}'_{l})$ and $q'_{l}(1)\neq q_{l}(p(\boldsymbol{q}'_{l})-1)$. This is described in Fig.~\ref{case4} where $d = \tilde{p}(\boldsymbol{q}'_{l})$ and $r=p(\boldsymbol{q}'_{l})-1$.
        According to the corresponding condition, there must exist a unique $k$ $(k \in {2,\cdots,r})$ such that $q'_{l}(k)=q'_{l}(1+r)$.
        Then similar to Case $2$ in the \emph{Basic System}, one can know that the AoI of the latest arrived update is greater that of the $k$-th latest update, which means that the AoI of the monitor would not be refreshed.
\end{itemize}

With the knowledge of the change of $\boldsymbol{x}$, $\boldsymbol{A}_{l}$ can be get explicitly and the change of $\boldsymbol{v}_{\boldsymbol{q}l}$ can be further deduced as shown in Table \ref{Tab_l_n_1stream}.

Then, for the differential equations used to describe the evolution of $\boldsymbol{x}$, we assert that it is the same as \eqref{nbq} since $\boldsymbol{x}$ still represents the age of the monitor and those of $N$ type-2 devices.

Next, to get $\boldsymbol{v}_{\boldsymbol{q}}$, it remains to deduce the steady state probability vector $\overline{\pi}_{\boldsymbol{q}}$ $(\boldsymbol{q}\in \mathcal{Q})$.
Since for any $\boldsymbol{q}\in \mathcal{Q}$, there are $(N+1)$ outgoing transitions whose transition rates are $u_{1}$ to
$u_{(N+1)}$ respectively, it has
\begin{equation}\label{sum_lambda_l2}
 \sum_{l\in\mathcal{L}_{\boldsymbol{q}}}\lambda^{(l)}=\sum_{i=1}^{N+1}u_{i}=\sum_{i=1}^{N}\mu_{i}+\lambda=\mu+\lambda.
\end{equation}
With \eqref{lambda_l2} and \eqref{sum_lambda_l2}, \eqref{steady state probability1} can be simplified as
\begin{equation}\label{steady state probability1_sim2}
  \overline{\pi}_{\boldsymbol{q}}=\frac{u_{q(1)}}{\mu+\lambda}\sum_{l\in\mathcal{L}_{\boldsymbol{q}}}\overline{\pi}_{\boldsymbol{q}_{l}}.
\end{equation}
Then one can get $\overline{\pi}_{\boldsymbol{q}}$ as follows
\begin{align}\label{n_1stream_steady state probability}
\overline{\pi}_{\boldsymbol{q}}&
             =\frac{u_{q(1)}}{\mu+\lambda}\sum_{\boldsymbol{q}'\in \mathcal{Q}_{\boldsymbol{q}}^{(1)}}\overline{\pi}_{\boldsymbol{q}'}
             \stackrel{\text{(i)}}{=}\frac{U(\boldsymbol{q})}{(\mu+\lambda)^{N}}\sum_{\boldsymbol{q}'\in\mathcal{Q}_{\boldsymbol{q}}^{(N)}}\overline{\pi}_{\boldsymbol{q}'}
                                       \notag\\
             &
             =\frac{U(\boldsymbol{q})}{(\mu+\lambda)^{N}}\sum_{\boldsymbol{q}'\in\mathcal{Q}}\overline{\pi}_{\boldsymbol{q}'}
             \stackrel{\text{(j)}}{=}\frac{U(\boldsymbol{q})}{(\mu+\lambda)^{N}},
\end{align}
where ${\text{(i)}}$ follows from expanding $\overline{\pi}_{\boldsymbol{q}'}$ by \eqref{steady state probability1_sim2} constantly, and ${\text{(j)}}$ follows from \eqref{steady state probability2}.

Since what we indeed concern about is $\overline{v}_{\boldsymbol{q}0}$, it just need to derive $\overline{v}'_{\boldsymbol{q}_{l}0}$($l\in \mathcal{L}'_{\boldsymbol{q}}$) which $\overline{v}_{\boldsymbol{q}0}$ is related to through the following equation that is obtained from \eqref{correlation vector} using \eqref{lambda_l2}, \eqref{nbq}, \eqref{sum_lambda_l2}, and \eqref{n_1stream_steady state probability}. 
\begin{equation}\label{stable vqi_sim2}
  \overline{v}_{\boldsymbol{q}i} = \frac{U(\boldsymbol{q})}{(\mu+\lambda)^{N+1}}+\frac{u_{q(1)}}{\mu+\lambda}\sum_{l\in \mathcal{L}'_{\boldsymbol{q}}}\overline{v}'_{\boldsymbol{q}_{l}i}, \boldsymbol{q}\in\mathcal{Q},i\in \{0\}\cup\mathcal{N}.
\end{equation}

Similar to the analysis in the \emph{Basic System}, if one can know $\overline{v}_{\boldsymbol{q}_li}$ $(l\in{\mathcal{L}^\prime_{\boldsymbol{q}}},~i\in \mathcal{N}_{\boldsymbol{q}_l}$) for a given $\boldsymbol{q}$, then one can get $\overline{v}_{\boldsymbol{q}0}$ by recursion. Then noticing that it requires to know $\overline{v}_{\boldsymbol{q}0}$ for $\boldsymbol{q}\in\mathcal{Q}$, it motivates to derive $\overline{v}_{\boldsymbol{q}i}$ for all $\boldsymbol{q}\in\mathcal{Q}$ and $i\in \mathcal{N}_{\boldsymbol{q}}$.
Let us derive these $\overline{v}_{\boldsymbol{q}i}$ first.
According to the value of $p_{i}(\boldsymbol{q})~(i\in \mathcal{N}_{\boldsymbol{q}})$, one can deduce the considered $\overline{v}_{\boldsymbol{q}i}$ as follows.

\begin{itemize}
\item $p_{i}(\boldsymbol{q})=1$: In this case, one can find that $\overline{v}'_{\boldsymbol{q}_{l}i}=0$ for all $l\in\mathcal{L}'_{\boldsymbol{q}}$ since $i=q(1)=q'_{l}(1)$ (cf. Tab. \ref{Tab_l_n_1stream}). Then one can get $\overline{v}_{\boldsymbol{q}i}$ from \eqref{stable vqi_sim2} as
    \begin{equation}\label{vqi_1_n1}
      \overline{v}_{\boldsymbol{q}i}=\frac{U(\boldsymbol{q})}{(\mu+\lambda)^{N+1}}.
    \end{equation}
\item  $2\le p_{i}(\boldsymbol{q})\le N$: In this case, one can find that $\overline{v}'_{\boldsymbol{q}_{l}i}=\overline{v}_{\boldsymbol{q}_{l}i}$ for all $l\in\mathcal{L}'_{\boldsymbol{q}}$ since $i\neq q'_{l}(1)$ and $i\neq 0$ (cf. Tab. \ref{Tab_l_n_1stream}). Then from \eqref{stable vqi_sim2}, it has that
    \begin{equation}\label{vqi2_n1}
    \overline{v}_{\boldsymbol{q}i}=\frac{U(\boldsymbol{q})}{(\mu+\lambda)^{N+1}}+\frac{u_{q(1)}}{\mu+\lambda}\sum_{l\in \mathcal{L}'_{\boldsymbol{q}}}\overline{v}_{\boldsymbol{q}_{l}i}.
    \end{equation}
It can be found that $p_{i}(\boldsymbol{q}_{l})=p_{i}(\boldsymbol{q})-1$ $(l\in \mathcal{L}'_{\boldsymbol{q}})$ from \eqref{ql_to_ql'}. Then one can get $\overline{v}_{\boldsymbol{q}i}$ by recursion as follows
\begin{align}\label{vqi_n_n1}
\overline{v}_{\boldsymbol{q}i}={}&\frac{U(\boldsymbol{q})}{(\mu+\lambda)^{N+1}}\notag\\&+\frac{u_{q(1)}}{\mu+\lambda}\sum_ {\tilde{\boldsymbol{q}}\in \mathcal{Q}_{\boldsymbol{q}}^{(1)}}\Bigl(\frac{U(\tilde{\boldsymbol{q}})} {(\mu+\lambda)^{N+1}}
                              +\frac{u_{\tilde{q}(1)}}{\mu+\lambda}\sum_{l\in\mathcal{L}'_{\tilde{\boldsymbol{q}}}}\overline{v}_{\tilde{\boldsymbol{q}}_{l}i}\!\Bigl)\notag\\
                              ={}&\frac{2U(\boldsymbol{q})}{(\mu+\lambda)^{N+1}}+\frac{u_{q(1)}u_{q(2)}}{(\mu+\lambda)^{2}}\sum_ {\tilde{\boldsymbol{q}}\in \mathcal{Q}_{\boldsymbol{q}}^{(2)}}\overline{v}_{\tilde{\boldsymbol{q}}i}\notag\\
                             \stackrel{\text{(k)}}{=}{}&\frac{(p_{i}(\boldsymbol{q})-1)U(\boldsymbol{q})}{(\mu+\lambda)^{N+1}}
                             \notag\\&+\frac{\prod_{j=1}^{p_{i}(\boldsymbol{q})-1}u_{q(j)}}{(\mu+\lambda)^{(p_{i}(\boldsymbol{q})-1)}} \frac{(\mu+\lambda)^{p_{i}(\boldsymbol{q})-1}U(\boldsymbol{q})}{\prod_{j=1}^{p_{i}(\boldsymbol{q})-1}u_{q(j)}(\mu+\lambda)^{N+1}}\notag\\
                             ={}&\frac{p_{i}(\boldsymbol{q})U(\boldsymbol{q})}{(\mu+\lambda)^{N+1}},
\end{align}
where ${\text{(k)}}$ follows from \eqref{vqi_1_n1} and \eqref{vqi2_n1}.
\end{itemize}

Combining \eqref{vqi_1_n1} and \eqref{vqi_n_n1}, for any $\boldsymbol{q}\in \mathcal{Q},i\in \mathcal{N}_{\boldsymbol{q}}$, it has
\begin{equation}\label{vqn_n1}
\overline{v}_{\boldsymbol{q}i}=\frac{p_{i}(\boldsymbol{q})U(\boldsymbol{q})}{(\mu+\lambda)^{N+1}}.
\end{equation}

Finally, according to the value of $p(\boldsymbol{q})$, $\tilde{p}(\boldsymbol{q})$ and whether device $q(1)$ is the source of the update repeats in the previous updates for the first time, one can deduce $\overline{v}_{\boldsymbol{q}0}$ as follows.
\begin{itemize}
  \item $\tilde{p}(\boldsymbol{q})=1$: In this case, one can find that $\overline{v}'_{\boldsymbol{q}_{l}0}=0$ for all $l\in\mathcal{L}'_{\boldsymbol{q}}$ (cf. Tab. \ref{Tab_l_n_1stream}).
      Then from \eqref{stable vqi_sim2}, it has
  \begin{align}\label{vq0_1_n1}
  \overline{v}_{\boldsymbol{q}0}&=\frac{U(\boldsymbol{q})}{(\mu+\lambda)^{N+1}}.
  \end{align}
  \item $1<\tilde{p}(\boldsymbol{q})<p(\boldsymbol{q})$: In this case, one can find that $\overline{v}'_{\boldsymbol{q}_{l}0}=\overline{v}_{\boldsymbol{q}_{l}0}$ for all $l\in\mathcal{L}'_{\boldsymbol{q}}$ (cf. Tab. \ref{Tab_l_n_1stream}). Based on \eqref{stable vqi_sim2}, it has
      \begin{align}\label{vq02_n1}
        \overline{v}_{\boldsymbol{q}0}  &=
                       \frac{U(\boldsymbol{q})}{(\mu+\lambda)^{(N+1)}}+\frac{u_{q(1)}}{\mu+\lambda}\sum_{l\in
                       \mathcal{L}'_{\boldsymbol{q}}}\overline{v}_{\boldsymbol{q}_{l}0}.
      \end{align}
  It also can be found that $\tilde{p}(\boldsymbol{q}_{l})=\tilde{p}(\boldsymbol{q})-1$ $(l\in\mathcal{L}'_{\boldsymbol{q}})$ from
  \eqref{ql_to_ql'} and $\tilde{p}(\boldsymbol{q})<p(\boldsymbol{q})$ is still satisfied.
  Then, the recursive method can be also used to deduce
  $\overline{v}_{\boldsymbol{q}0}$ as follows
\begin{align}\label{vq0_2_n1}
\overline{v}_{\boldsymbol{q}0}={}&\frac{U(\boldsymbol{q})}{(\mu+\lambda)^{N+1}}\notag\\&+\!\frac{u_{q(1)}}{\mu+\lambda}\sum_ {\tilde{\boldsymbol{q}}\in \mathcal{Q}_{\boldsymbol{q}}^{(1)}}\Bigl(\frac{U(\tilde{\boldsymbol{q}})} {(\mu+\lambda)^{N+1}}
                              +\frac{u_{\tilde{q}(1)}}{\mu\!+\!\lambda}\sum_{l\in\mathcal{L}'_{\tilde{\boldsymbol{q}}}}\overline{v}_{\tilde{\boldsymbol{q}}_{l}0}\!\Bigl)\notag\\
                              ={}&\frac{2U(\boldsymbol{q})}{(\mu+\lambda)^{N+1}}+\frac{u_{q(1)}u_{q(2)}}{(\mu+\lambda)^{2}}\sum_ {\tilde{\boldsymbol{q}}\in \mathcal{Q}_{\boldsymbol{q}}^{(2)}}\overline{v}_{\tilde{\boldsymbol{q}}0}\notag\\
                             \stackrel{\text{(l)}}{=}{}&\frac{(\tilde{p}(\boldsymbol{q})-1)U(\boldsymbol{q})}{(\mu+\lambda)^{N+1}}
                             \notag\\&+\frac{\prod_{j=1}^{\tilde{p}(\boldsymbol{q})-1}u_{q(j)}}{(\mu+\lambda)^{(\tilde{p}(\boldsymbol{q})-1)}} \frac{(\mu+\lambda)^{\tilde{p}(\boldsymbol{q})-1}U(\boldsymbol{q})}{\prod_{j=1}^{\tilde{p}(\boldsymbol{q})-1}u_{q(j)}(\mu+\lambda)^{N+1}}\notag\\
                             ={}&\frac{\tilde{p}(\boldsymbol{q})U(\boldsymbol{q})}{(\mu+\lambda)^{N+1}},
\end{align}
where \text{(l)} follows from \eqref{vq0_1_n1} and \eqref{vq02_n1}.

Combining \eqref{vq0_1_n1} and \eqref{vq0_2_n1}, we have
\begin{equation}\label{vq0_3_n1}
  \overline{v}_{\boldsymbol{q}0}=\frac{\tilde{p}(\boldsymbol{q})U(\boldsymbol{q})}{(\mu+\lambda)^{N+1}},\quad \tilde{p}(\boldsymbol{q})<p(\boldsymbol{q}).
\end{equation}
  \item $2 \leq p(\boldsymbol{q}) < \tilde{p}(\boldsymbol{q})$: This case is very similar to Case 1 of deducing $\overline{v}_{\boldsymbol{q}0}$ in the \emph{Basic System}. Hence, here we omit the derivation and only give the expression as follows
  \begin{equation}\label{vq0_4_n1}
  \overline{v}_{\boldsymbol{q}0}=\frac{p(\boldsymbol{q})U(\boldsymbol{q})}{(\mu+\lambda)^{N+1}}, \quad \tilde{p}(\boldsymbol{q})>p(\boldsymbol{q}).
  \end{equation}
  \item $\tilde{p}(\boldsymbol{q})=p(\boldsymbol{q})=N+1$: In this case, one can find that $\overline{v}'_{\boldsymbol{q}_{l}0}$ is equal to $\overline{v}_{\boldsymbol{q}_{l}q(1)}$ when $q_{l}(N)=q(1)$ and $\overline{v}_{\boldsymbol{q}_{l}0}$ otherwise (cf. Tab. \ref{Tab_l_n_1stream}).
There is only one transition $l$ such that $q_{l}(N)=q(1)$. Let $l'$ denote this transition, i.e., $q_{l'}(N)=q(1)$.
Then one can find that $p_{q(1)}(\boldsymbol{q}_{l'})=N$.
For the rest $l$, it has
        \begin{equation}\label{}
        \begin{cases}
           p(\boldsymbol{q}_{l})=N<\tilde{p}(\boldsymbol{q}_{l})=N+1, &\!\!\! q_{l}(N)\ne N+1,\\
           \tilde{p}(\boldsymbol{q}_{l})=N<p(\boldsymbol{q}_{l})=N+1, &\!\!\! q_{l}(N)=N+1.
        \end{cases}
      \end{equation}
It can be found that $\min\{\tilde{p}(\boldsymbol{q}_{l}),p(\boldsymbol{q}_{l})\}$ always equals $N$.
        Thus, from \eqref{vq0_3_n1} and \eqref{vq0_4_n1}, it has
      \begin{equation}\label{vq0_5_n1}
        \overline{v}_{\boldsymbol{q}_{l}0}=\frac{NU(\boldsymbol{q})u_{q_{l}(N)}}
                                            {u_{q(1)}(\mu+\lambda)^{N+1}},\quad l\in\mathcal{L}'_{\boldsymbol{q}},~l\neq l'.
      \end{equation}
Then one can get $\overline{v}_{\boldsymbol{q}0}$ as follows
      \begin{align}\label{vq0_6_n1}
        \overline{v}_{\boldsymbol{q}0}={} & \frac{U(\boldsymbol{q})}{(\mu+\lambda)^{N+1}}+\frac{u_{q(1)}}{\mu+\lambda}\sum_{\substack{l\in \mathcal{L}'_{\boldsymbol{q}},l\ne l'}}\overline{v}_{\boldsymbol{q}_{l}0}
        \notag\\ &+
                                            \frac{u_{q(1)}}{\mu+\lambda}\overline{v}_{\boldsymbol{q}_{l'}q(1)}  \notag\\
                                      \stackrel{\text{(m)}}{=}{} & \frac{U(\boldsymbol{q})}{(\mu+\lambda)^{N+1}}+\frac{u_{q(1)}}{\mu+\lambda}\frac{NU(\boldsymbol{q})/u_{q(1)}}
                                            {(\mu+\lambda)^{N+1}}(\mu+\lambda)
                                            \notag\\
                                           &+\frac{u_{q(1)}}{\mu+\lambda}\frac{NU(\boldsymbol{q})}{(\mu+\lambda)^{N+1}}(1-\frac{u_{q_{l'}(N)}}{u_{q(1)}})
                                            \notag\\
                                      ={} & \frac{(N+1)U(\boldsymbol{q})}{(\mu+\lambda)^{N+1}}.
      \end{align}
where $\text{(m)}$ follows from \eqref{vq0_5_n1} and \eqref{vqn_n1}.
\end{itemize}
Combining \eqref{vq0_3_n1}, \eqref{vq0_4_n1}, and \eqref{vq0_6_n1}, for all $\boldsymbol{q}\in \mathcal{Q}$, it has
\begin{equation}\label{vq0_n1}
  \overline{v}_{\boldsymbol{q}0}=\frac{\min\{p(\boldsymbol{q}),\tilde{p}(\boldsymbol{q})\}U(\boldsymbol{q})}{(\mu+\lambda)^{N+1}}.
\end{equation}
Further using \eqref{vq0_n1} and \eqref{average AoI}, one can get $\Delta_{N,M}^{\text{Hybrid}}$ as \eqref{n_1streamaverageAoI}.

\subsection{Proof of Corollary 2}\label{Corollary_2_proof}
According to whether there are updates from the equivalent type-2 device in the previous $N$ updates, we divide the discrete state set $\mathcal{Q}$ into two disjoint subsets: $\mathcal{Q}_{\text{w/o}}$ and $\mathcal{Q}_{\text{w}}$. In particular,
$\mathcal{Q}_{\text{w}}$ and $\mathcal{Q}_{\text{w/o}}$ are the subset of states with and without update that come from the type-2 device in the previous $N$ updates, respectively.
Thus, for $\boldsymbol{q}\in \mathcal{Q}_{\text{w/o}}$, it has $\min\{p(\boldsymbol{q}),\tilde{p}(\boldsymbol{q})\} = p(\boldsymbol{q})$ and $U(\boldsymbol{q}) = M(\boldsymbol{q})$.
Then, $\Delta_{N,M}^{\text{Hyb-Homo}}$ can be expressed as
\begin{align}\label{delta_hyb_homo}
  \Delta_{N,M}^{\text{Hyb-Homo}}={} & \sum_{\boldsymbol{q}\in\mathcal{Q}_{\text{w/o}}}\frac{p(\boldsymbol{q})M(\boldsymbol{q})}{(\mu+\lambda)^{(N+1)}} \notag\\
  &+ \sum_{\boldsymbol{q}\in\mathcal{Q}_{\text{w}}}\frac{\min\{p(\boldsymbol{q}),\tilde{p}(\boldsymbol{q})\}U(\boldsymbol{q})}{(\mu+\lambda)^{(N+1)}}.
\end{align}

Let us denote the first and the second terms in \eqref{delta_hyb_homo} as $\Delta_{\text{w/o}}$ and $\Delta_{\text{w}}$, respectively.
Note that $\mathcal{Q}_{\text{w/o}}$ is the same with the discrete state set in the \emph{Basic System}, then $\Delta_{\text{w/o}}$ can be derived
\begin{align}\label{delta_without}
  \Delta_{\text{w/o}} & = \sum_{\boldsymbol{q}\in\mathcal{Q}_{\text{w/o}}}\frac{p(\boldsymbol{q})M(\boldsymbol{q})}{\mu^{(N+1)}}\frac{\mu^{(N+1)}}{(\mu+\lambda)^{(N+1)}}\notag\\
   & =
  \frac{1}{\mu_{i}}\sum\limits_{r = 1}^{N } \frac{\binom{N-1}{r-1} (r+1)!}{N^{r+1}}\left(\frac{\mu}{\mu+\lambda}\right)^{N+1}.
\end{align}

For $\Delta_{\text{w}}$, we use an approach similar to that adopted in the proof of corollary 1, which yields
\begin{align}\label{}
  \Delta_{\text{w}} = {}&\sum_{k = 1}^{N}\left(\frac{C(k)p(\boldsymbol{q})\mu_{i}^{N-k}\lambda^{k}}{(\mu+\lambda)^{(N+1)}} +
  \frac{\tilde{C}(k)\tilde{p}(\boldsymbol{q})\mu_{i}^{N-k}\lambda^{k}}{(\mu+\lambda)^{(N+1)}}\right),\notag\\
  ={}& A(k) + B(k),
\end{align}
where $C(k)$ and $\tilde{C}(k)$ are the number of states that satisfy $\min\{p(\boldsymbol{q}),\tilde{p}(\boldsymbol{q})\}= p(\boldsymbol{q})$ and $\min\{p(\boldsymbol{q}),\tilde{p}(\boldsymbol{q})\}=\tilde{p}(\boldsymbol{q})$ respectively when there are $k$ updates from the type-2 device in the previous $N$ updates.

When $\min\{p(\boldsymbol{q}),\tilde{p}(\boldsymbol{q})\}= p(\boldsymbol{q})$, it has $p(\boldsymbol{q}) < \tilde{p}(\boldsymbol{q})$.
Then, the range of $p(\boldsymbol{q})$ is $2 \leq p(\boldsymbol{q}) \leq N-k$ which implies that $N-k-1>0$, i.e., $N-k \geq 2$.
For a fixed $p(\boldsymbol{q}) = p$, there are $k$ updates from the type-2 device from the $(1 + p)$-th latest to the $N$-th latest updates and the rest $N-p-k$ updates can come from any type-1 device. Moreover, the previous $p-1$ updates (from the latest to the $p-1$-th latest) must come from different type-1 devices. In particular, the source of the $p$-th latest update should be the same as one among the previous $p-1$ updates.
Accordingly, it has

\begin{align}\label{}
  A(k) & = \sum_{p=2}^{N-k}\frac{\binom{N-p}{k}N^{N-p-k}N\binom{N-1}{p-2}(p-1)!p\mu^{N-k}\lambda^{k}}{N^{N-k}(\mu+\lambda)^{N+1}}\notag\\
   & = \sum_{p=2}^{N-k}\frac{\binom{N-p}{k}\binom{N-1}{p-2}p!\mu^{N-k}\lambda^{k}}{N^{p-1}(\mu+\lambda)^{N+1}}.
\end{align}

When $\min\{p(\boldsymbol{q}),\tilde{p}(\boldsymbol{q})\}= \tilde{p}(\boldsymbol{q})$, it has $\tilde{p}(\boldsymbol{q}) < p(\boldsymbol{q})$.
Then, the range of $\tilde{p}(\boldsymbol{q})$ is $1 \leq \tilde{p}(\boldsymbol{q}) < N - k + 1$ which always holds for $\mathcal{Q}_{\text{w}}$.
Similarly, one can get $B(k)$ as
\begin{align}\label{}
  B(k) & =\sum_{p=1}^{N-k+1}\frac{\binom{N-p}{k-1}N^{N-p-k+1}\binom{N}{p-1}(p-1)!p\mu^{N-k}\lambda^{k}}{N^{N-k}(\mu+\lambda)^{N+1}}\notag \\
   & =\sum_{p=1}^{N-k+1}\frac{\binom{N-p}{k-1}\binom{N}{p-1}p!\mu^{N-k}\lambda^{k}}{N^{p-1}(\mu+\lambda)^{N+1}}.
\end{align}

Note that $A(k)$ exists when $N-k-1>0$, then it has
\begin{equation}\label{delta_with}
  \Delta_{\text{w}} = \sum_{k=1}^{N}(\mathbb{I}_{\{N-k-1>0\}}A(k)+B(k)).
\end{equation}

Plugging \eqref{delta_without} and \eqref{delta_with} into \eqref{delta_hyb_homo}, one can get $\Delta_{N,M}^{\text{Hyb-Homo}}$ as \eqref{collary2}.

\ifCLASSOPTIONcaptionsoff
  \newpage
\fi

\bibliographystyle{IEEEtran}
\bibliography{mybib}

\end{document}